\documentclass[10pt, twocolumn]{IEEEtran}

\ifCLASSINFOpdf
  \usepackage[pdftex]{graphicx}
\else
\fi

\usepackage{algorithm}  
\usepackage{algpseudocode}
\usepackage{amsfonts,amsmath,amsthm,amssymb,graphicx,algorithm,epstopdf,cite,url,subfigure}
\usepackage{psfrag,amssymb,amsmath,pifont,cite,graphics,graphicx,epsfig,subfigure,amscd,helvet,multirow,enumerate}
\usepackage{lmodern}
\usepackage{color}
\usepackage{float}
\usepackage{epstopdf}
\usepackage{accents}
\usepackage{amssymb,amsmath,amscd,cite,enumerate}
\usepackage{arydshln}
\usepackage{graphics,graphicx,epsfig,psfrag,subfigure,threeparttable,url}
\usepackage[colorlinks,citecolor=black,linkcolor=black]{hyperref}
\usepackage{algpseudocode}
\usepackage{dblfloatfix}
\usepackage{blindtext}
\usepackage[english]{babel}
\newtheorem{theorem}{Theorem}
\newtheorem{lemma}{Lemma}
\newtheorem{definition}{Definition}
\newtheorem{proposition}{Proposition}

\begin{document}
%
\title{Preconditioned Multiple Orthogonal Least Squares and Applications in Ghost Imaging via Sparsity Constraint}


\author{Zhishen Tong, Jian Wang,~\IEEEmembership{Member,~IEEE} and Shensheng Han
 
\thanks{Z. Tong and S. Han are with the Key Laboratory for Quantum Optics and Center for Cold Atom Physics of CAS, Shanghai Institute of Optics and Fine Mechanics, Chinese Academy of Sciences, Shanghai, 201800, China. They are also with the Center of Materials Science and Optoelectronics Engineering, University of Chinese Academy of Sciences, Beijing 100049, China.  (e-mail: tongzhishen@siom.ac.cn; sshan@mail.shcnc.ac.cn)

J.~Wang is with the School of Data Science, Fudan University, Shanghai 200433, China.  (e-mail: jian\_wang@fudan.edu.cn) (Corresponding author: Jian Wang)
}
}

\markboth{}
{Shell \MakeLowercase{\textit{et al.}}: Bare Demo of IEEEtran.cls for Journals}
%




\IEEEtitleabstractindextext{%
\begin{abstract}
Ghost imaging via sparsity constraint (GISC) can recover  objects from the intensity fluctuation of light fields even when the sampling rate is far below the Nyquist sampling rate. 
In this paper, we develop an efficient algorithm called the preconditioned multiple orthogonal least squares (PmOLS) for solving the GISC reconstruction problem. 
%
%
Our analysis shows that the PmOLS algorithm perfectly recovers any $n$-dimensional $K$-sparse signal from  $m$ random linear samples of the signal with probability exceeding
$
 1 - 3 n^2   e^{ - c m / K^2 }. \nonumber
$
Simulations and experiments demonstrate that the proposed algorithm has very competitive imaging quality compared to the state-of-the-art methods. 
%
\end{abstract}
\begin{IEEEkeywords}
Ghost imaging (GI),  sparsity, compressive sensing (CS), preconditioning, multiple orthogonal least squares (mOLS).

\end{IEEEkeywords}
}

\maketitle
 

\IEEEdisplaynontitleabstractindextext

%
\IEEEpeerreviewmaketitle


\section{Introduction}\label{sec:intro}
\subsection{Ghost Imaging via Sparsity Constraint}
 \IEEEPARstart As a novel imaging technique, ghost imaging (GI) extracts objects by exploiting the intensity fluctuation characteristics of light fields, which has received considerable attention in the last few decades~\cite{ Pittman1995twophotonquantum,
 Bennink2004quantumclassical, 
 Chenjing2004incoherentcoincidence,
 Scarcelli2006twophotoncorrelationandcorrelationIntensityfluctuations}. In a nutshell, GI consists of two major processes: i) sampling and ii) reconstruction. In the sampling process, objects are encoded by the random intensity fluctuation of light fields. Whereas in the reconstruction process, objects are decoded from the second-order correlation of intensity fluctuation. Applications of GI include Lidar staring imaging~\cite{Zhao2012Lidar}, single-shot multi-spectral imaging~\cite{Sahoo2017Spectralcamera},  X-ray Fourier transform diffraction imaging~\cite{Chenjing2004incoherentcoincidence}, etc.

Although the GI technique has demonstrated advantages in complex imaging environments~\cite{Ferri2010DGI}, 
its imaging capacity is fundamentally limited to the number of samples.  In general, more samples of objects lead to  better imaging quality. However, a large number of samples is often associated with a high acquisition cost, which in turn limits the practical applications of GI. Recently, therefore, much effort has been devoted to improve the imaging quality of GI based on the given number of samples; See, for examples, the differential GI (DGI)~\cite{Ferri2010DGI}, normalized GI (NGI)~\cite{Sun2012NGI}, and pseudo-inverse for GI (PGI)~\cite{Zhang2014PGI}. While these methods have achieved more or less improvement of the imaging quality, they are still demanding in the number of samples.

Inspired by the prevalent compressive sensing~(CS) theory~\cite{donoho2006compressed,candes2005decoding,
candes2008restricted}, which exploits sparsity of the imaging objects, GI via sparsity constraints (GISC) has recently been proposed~\cite{Gong2012superresolutionGISC}. In general, the reconstruction problem of GISC is given by
\begin{equation}
\mathop {\min }\limits_\mathbf{x} {\left\| \mathbf{x} \right\|_0}~~ \rm{subject~ to}~~~~ \mathbf{y}_0 = \mathbf{\Psi x},
\label{equation: measurement}
\end{equation}
where $\mathbf{y}_0 \in {\mathbb{R} ^{m}}$ is the sampling signals measured by a bucket detector, $\mathbf{x} \in {\mathbb{R} ^{n}}$ represents the object, which is typically $K$-sparse (i.e, $\|\mathbf{x}\|_0 \leq K$ where $\left\| \hspace{.5mm} \cdot \hspace{.5mm} \right\|_0$ denotes the $\ell_0$-norm), and $\mathbf{\Psi} \in {\mathbb{R} ^{m \times n}}$ is the sampling matrix consisting of the light intensity recorded by a pixelated detector ($\mathbf{\Psi}_{ij} \geq 0$, $\forall i,j$).  
The key advantage of the GISC technique lies in that it works even when the sampling rate is far below the Nyquist sampling rate~\cite{Bromberg2009ghostimagingwithasingledetector,Katz2009compressiveghostimaging}. Owing to this advantage, GISC has greatly promoted the industrial applications of GI, such as the super-resolution imaging~\cite{Gong2012superresolutionGISC}, three-dimensional (3-D) computational imaging with single-pixel detectors~\cite{sun20133dcomputationalimagingsinglepiexldetectors} and single-shot spectral camera~\cite{Liu2015Spectralcamera}.

 \subsection{Preconditioning}
There has been much evidence that the imaging quality of GISC depends heavily on the property of the sampling matrix. In particular, GISC produces faithful imaging quality when the sampling matrix has sufficiently small mutual coherence, which is defined as~\cite{donoho2006compressed}
    \begin{equation}
    \mu (\mathbf{\Psi}) = \mathop {\max}\limits_{1 \leq i < j \leq n} \frac{  | \langle \psi_i, \psi_j \rangle | } { \| \psi_i \|_2  \| \psi_j \|_2 },
    \label{mutual coherence}
    \end{equation}
where $\| \hspace{.5mm} \cdot \hspace{.5mm} \|_2$ is the $\ell_2$-norm and $\psi_i$ denotes the $i$-th column of the sampling matrix $\bf{\Psi}$. For a given imaging system, however, the mutual coherence $\mu ({\bf{\Psi}})$ of the sampling matrix is physically determined and could be expensive or hard to reduce in hardware. Thus, much attention has been focused on refining $\mu ({\bf{\Psi}})$ via post processing. Typically, by multiplying a pre-designed matrix $\mathbf{P}$ to the sampling signals $\mathbf{y}_0$, problem~\eqref{equation: measurement} can be converted to
  \begin{equation}
\mathop {\min }\limits_\mathbf{x} {\left\| \mathbf{x} \right\|_0}~~ \rm{subject~ to}~~~~ \mathbf{P}  \mathbf{y}_0  = \mathbf{P} \mathbf{\Psi}  \mathbf{x}.
\label{equation: preconditin_measurement}
\end{equation}
We would like the new sampling matrix $\mathbf{P} \mathbf{\Psi}$ to have smaller mutual coherence than $\mathbf{\Psi}$ does, so that the underlying signal $\mathbf{x}$ could be reconstructed from $\mathbf{P} \mathbf{y}_0$ with hopefully better quality. In numerical linear algebra, the method of multiplying a pre-designed matrix to the samples  $\mathbf{y}_0$ so as to improve the recovery quality of vector $\mathbf{x}$ is often referred to as preconditioning~\cite{Benzi2002preconditioning}, where matrix $\rm{\mathbf{P}}$ is called the preconditioner.

    In CS, there have been many studies on preconditioning. For example, Elad~\cite{Elad2007optimizedprojection} proposed a shrinkage operation on the off-diagonal entries of the Gram matrix of $\mathbf{P} \mathbf{\Psi}$ to refine its average mutual coherence. Duarte-Carvajalino {\it et al.}~\cite{duarte2009Sparsifyingdictionaryoptimization} suggested a matrix optimization method to optimize the preconditioner $\mathbf{P}$ for the given sampling matrix $\mathbf{\Psi}$, which makes the Gram matrix of $\mathbf{P} \mathbf{\Psi}$ approach an identity matrix.   
    Based on the frame theory, Tsiligianni {\it et al.}~\cite{Tsiligianni2015preconditionofIUNTF} iteratively constructed the matrix $\mathbf{P} \mathbf{\Psi}$ towards a unit norm tight frame (UNTF). 
    While these preconditioning methods have achieved much improvement on the reconstruction quality, they have limitations in both theory and practice. To be specific, the preconditioners of these methods are all obtained in an iterative fashion, which may be computationally involved in practice. Moreover, the iterations unavoidably bring in difficulties to the theoretical analyses of the preconditioners. To date, little has been known about the theoretical guarantees of these existing preconditioning methods~\cite{Elad2007optimizedprojection,duarte2009Sparsifyingdictionaryoptimization,Tsiligianni2015preconditionofIUNTF}. 

\begin{figure}[t]
\centering
{\hspace{-3mm}\includegraphics[width = 85mm, height = 85mm] {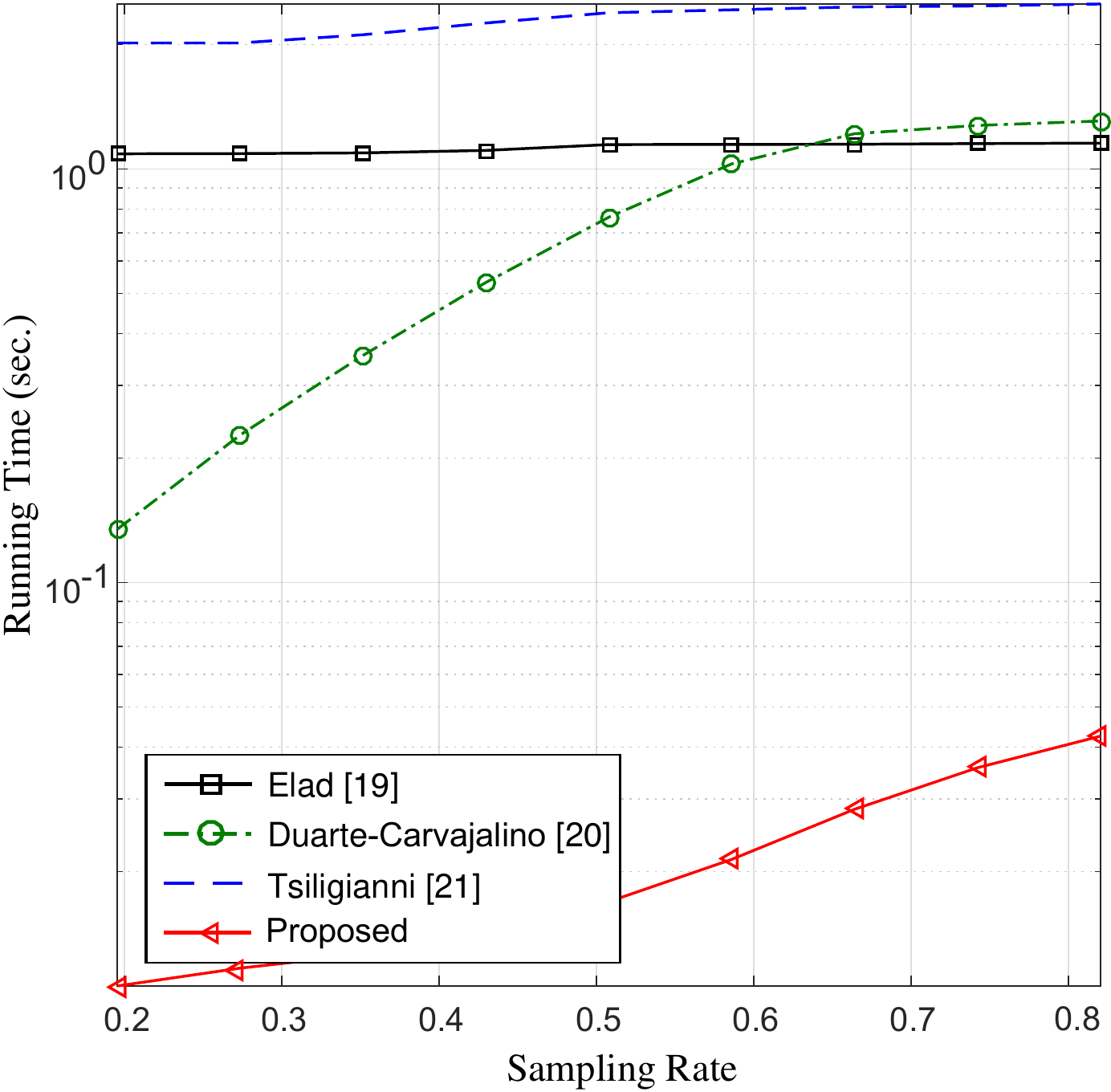}}
\caption{The empirical running time of different preconditioning methods as a function of sampling rate. }
\label{fig:running_time_comparsion} 
\end{figure}

     In this paper, we propose an efficient preconditioning method called the pseudo-inverse preconditioning (PIP) to refine the mutual coherence of sampling matrices. We consider the typical setting of GISC where down-sampling is used (i.e., $m < n$). Our preconditioner $\mathbf{P}$ is literally the pseudo-inverse matrix of the original sampling matrix $\mathbf{\Psi}$,  
     \begin{equation}
      \mathbf{P} := \mathbf{\Psi}^T \left( \mathbf{\Psi} \mathbf{\Psi}^T \right) ^{-1}.  \label{eq:eq4}
      \end{equation} 
In the following, we  summarize the advantages of the proposed PIP method in two aspects. 

\begin{enumerate}[i)]
\item {\bf Speed}: The empirical running time of different preconditioning methods as a function of sampling rate is plotted in Fig.~\ref{fig:running_time_comparsion}. In our simulation, we consider random Gaussian sampling matrices with $n = 256$ and entries drawn {\it i.i.d.} from Gaussian distribution $\mathcal{N}(0, {1}/{m})$ for various sampling rates. For all the aforementioned preconditioning methods, the running time was measured using the MATLAB program under the $2$-core $2.6$GHz
$64$-bit processor, $256$Gb RAM and Windows Server $2012$
R$2$ environments. For comparative purpose, the number of iterations for these iterative methods~\cite{Elad2007optimizedprojection,duarte2009Sparsifyingdictionaryoptimization,Tsiligianni2015preconditionofIUNTF} is uniformly set to ten. It can be observed that the proposed PIP method is computationally much more efficient than the iterative ones. In particular, the running time of PIP is less than $10$\% of the other  methods for the whole region of sampling rate. 

\vspace{1mm}

\item {\bf Performance guarantee}:  In the CS literature, random Gaussian sampling matrices have been extensively studied as they often lead to promising recovery performance as well as elegant guarantees (see, e.g.,~\cite{donoho2006compressed,
candes2005decoding}). In this paper, we also consider the random Gaussian sampling matrices to study the effectiveness of the proposed PIP method.  In particular, our analysis shows that the mutual coherence $\mu(\mathbf{P} \mathbf{\Psi})$  of the preconditioned sampling matrix is well bounded with high probability.  To the best of our knowledge, this result is the first theoretical guarantee on preconditioning. 
\end{enumerate}

\subsection{Reconstruction Methods for Solving GISC}

While preconditioning methods are useful for improving the recovery quality of signals, the performance of GISC also depends on the reconstruction method. Generally speaking, most of the algorithms currently used for solving the GISC problem are derived from CS, which can be roughly grouped into two categories: i) those relying on optimization and ii) those using greedy search. The optimization-based approaches replace the $\ell_0$-norm with $\ell_1$-norm and translate the non-convex problem in~\eqref{equation: measurement} into a convex problem: 
 \begin{equation}
\mathop {\min }\limits_\mathbf{x} {\left\| \mathbf{x} \right\|_1}~~ \rm{subject~ to}~~~~ \mathbf{y}_0 = \mathbf{\Psi x},
\label{equation: measurementl1}
\end{equation}
which is known as basis pursuit (BP)~\cite{chen2001atomic}. In the second category,  greedy algorithms have received considerable attention due to their computational simplicity and competitive performance. Representative methods include orthogonal matching pursuit (OMP)~\cite{Pati1993OMP,davenport2010analysis,liu2012orthogonal,mo2012remarks,wang2012Generalized,ding2013perturbation} and orthogonal least squares (OLS) \cite{Chen1989OLS,Soussen2013OLSsufficient,wen2017OLS}. They sequentially construct the support (i.e., the positions of nonzero elements) of sparse signals to minimize the representation distance to the samples $\mathbf{y}_0$.

As a natural extension of the OLS algorithm, multiple orthogonal least squares (mOLS) \cite{wang2017MOLS} selects multiple indices at a time corresponding to the $s$ ($\geq 1$) largest correlation magnitudes between the residual and the columns of sampling matrix. 
%
Thanks to the selection of multiple correct indices per iteration, the mOLS algorithm often converges much faster than the conventional OLS algorithm. While a massive reduction of the computation burden is achieved, the mOLS algorithm nevertheless generates promising reconstruction performance~\cite{wang2017MOLS}. Interested readers are referred to~\cite{wang2017MOLS} for more details about the mOLS algorithm. 

In this paper, we propose a preconditioned mOLS (PmOLS) algorithm for solving the GISC reconstruction problem. Basically, the PmOLS algorithm can be viewed as the mOLS algorithm with inputs being the preconditioned sampling matrix  and preconditioned samples; See Algorithm~\ref{tab:pmols} for a mathematical description of PmOLS. Particularly, we would like the preconditioned sampling matrix to have better mutual coherence property so that the PmOLS algorithm could reconstruct sparse signals with higher quality. Theoretical analysis shows that the PmOLS algorithm exactly solves the GISC problem with probability exceeding
\begin{equation}
 1 - 3 n^2   e^{ - c m / K^2 }. \nonumber
\end{equation}
Therefore,  the choice of $m \geq c K^2 \log (n/\epsilon)$ is sufficient to ensure the success probability exceeding $1 - \epsilon$.

\subsection{Organization}
The rest of this paper is organized as follows. In Section~\ref{sec:II}, we introduce some notations, definitions and lemmas that will be used in this paper. In Section~\ref{sec:III}, we introduce the PmOLS algorithm and analyze its theoretical performance in solving the GISC problem. In Section~\ref{sec:V}, we study the simulation and experimental performance of the PmOLS algorithm. In Section~\ref{sec:VI}, we discuss several interesting issues raised from our analysis and experiments. Concluding remarks are given in Section~\ref{sec:VII}.

\begin{algorithm}[t]  
  \caption{ The PmOLS Algorithm}  
   \label{tab:pmols} 
   \begin{algorithmic}[1]  
        \Require 
          sampling matrix $\mathbf{\Psi} \in \mathbb{R}^{m \times n}$,
                     samples $\mathbf{y}_0 \in \mathbb{R}^{m}$,
                     sparsity $K$, residual tolerant ${\bf tol}$
                      and number of indices in each selection $ s \leq \min \{K,  \frac{m}{K} \}$.
        \Ensure  
        estimated support set $\hat{T}$ and estimated signal $\hat{\mathbf{x}}$.
         \Function {PmOLS} {$ \mathbf{\Psi}, \mathbf{y}_0, K, {\bf tol}, s $}
                \State {\bf Initialization:}
                \State ~~~~preconditioner $ \mathbf{P} \gets \mathbf{\Psi}^{T} ( \mathbf{\Psi} \mathbf{\Psi}^{T} )^{-1}$;
                 \State ~~~~preconditioned samples
                $\mathbf{y} \gets \mathbf{P} \mathbf{y}_0$;
                \State ~~~~preconditioned sampling matrix $ \mathbf{\Phi} \gets \mathbf{P} \mathbf{\Psi}$;
                \State ~~~~iteration count $ k \gets 0 $;
                \State ~~~~estimated support $ \mathcal{S}^{0} \gets \emptyset $;
                \State ~~~~residual $ \mathbf{r}^{0} \gets \mathbf{y} $.
        \While{ $ \|\mathbf{r}^k\|_2 > {\bf tol} $ \text{and} $ k < \min \{ K,  \frac{m}{K} \}$}
        \State $ k = k + 1 $.
        \State Select $\mathcal{S}^{k} \hspace{-.5mm} = \hspace{-.25mm} \mathcal{S}^{k - 1} \cup \hspace{.25mm} \underset { \mathcal{S}: | \mathcal{S} | = s  } { \arg \min} \hspace{-.25mm}  \sum \nolimits_{ i \in \mathcal{S} }  \hspace{-.25mm} \| \mathcal{P}_{ \mathcal{S}^{ k - 1 } \cup \{ i \} } ^ \bot \mathbf{y} \|_2^2$; 
        \State Estimate $ \mathbf{x}^{k} = \underset{ \mathbf{u} : \text{supp} ( \mathbf{u} ) = \mathcal{S}^k} { \arg \min} \| \mathbf{y} - \mathbf{\Phi} \mathbf{u} \|_2$;
        \State Update $ \mathbf{r}^k = \mathbf{y} - \mathbf{\Phi} \mathbf{x}^k $.
        \EndWhile
        \State \Return{estimated support $\hat{T} =
         \underset{ \mathcal{S}: | \mathcal{S} | = K } { \arg \min } \| \mathbf{x}^k -
         \mathbf{x}^k_\mathcal{S} \|_2$, estimated signal $\hat{ \mathbf{x} }$ obeying $ \hat{ \mathbf{x} }_{ \Omega \setminus \hat{T} } = \mathbf{0} $ and $ \hat{ \mathbf{x} }_{ \hat{T} } = \mathbf{ \Phi }_{ \hat{T} }^\dag \mathbf{y}$.}
        \EndFunction
  \end{algorithmic}  
\end{algorithm}

\section{Preliminaries} \label{sec:II}
 \subsection{Notations and Definitions}
We first briefly explain some notations that will used throughout this paper. Let $\Omega = \{1,2,\cdots,n\}$ and let $T=\emph{supp}({\mathbf{x}})=\{i|i\in\Omega,x_i\neq 0\}$ denote the support of vector $\mathbf{x}$. For $\mathcal{S} \subseteq \Omega$, $|\mathcal{S}|$ is the cardinality of $\mathcal{S}$. $T \backslash \mathcal{S} = \{i|i\in T ~\text{but}~ i\notin \mathcal{S}\}$. 
$\mathbf{x}_{\mathcal{S}}\in \mathbb{R}^{|\mathcal{S}|}$ is the restriction of the vector $\mathbf{x}$ to the elements with indices in $\mathcal{S}$. $\mathbf{\Phi}_{\mathcal{S}}\in \mathbb{R}^{m\times|\mathcal{S}|}$ is a submatrix of $\mathbf{\Phi}$ that only contains columns indexed by $\mathcal{S}$. If $\mathbf{\Phi}_{\mathcal{S}}$ is full column rank, then $\mathbf{\Phi}_{\mathcal{S}}^\dagger = (\mathbf{\Phi}^{T}_{\mathcal{S}}\mathbf{\Phi}_{\mathcal{S}})^{-1}\mathbf{\Phi}^{T}_{\mathcal{S}}$ is the pseudoinverse of $\mathbf{\Phi}_{\mathcal{S}}$. $\emph{span}(\mathbf{\Phi}_{\mathcal{S}})$ is the span of columns in $\mathbf{\Phi}_{\mathcal{S}}$. $\mathcal{P}_{\mathcal{S}} = \mathbf{\Phi}_{\mathcal{S}}\mathbf{\Phi}_{\mathcal{S}}^\dagger$ is the projection onto $\emph{span}(\mathbf{\Phi}_{\mathcal{S}})$. $\mathcal{P}_{\mathcal{S}}^\perp = \mathbf{I}-\mathcal{P}_{\mathcal{S}}$ is the projection onto the orthogonal complement of $\emph{span}(\mathbf{\Phi}_{\mathcal{S}})$ where $\mathbf{I}$ is the identity matrix. 


  \begin{definition}(Restricted isometry property (RIP)~\cite{candes2005decoding})
  A matrix $\mathbf{\Phi}$ is said to satisfy the $RIP$ of order $K$ if there exists a constant $\delta \in [0,1)$ such that
  \begin{equation} \label{rip}
      (1-\delta)\|\mathbf{x}\|_2^2\le\|\mathbf{\Phi x}\|_2^2\le(1+\delta)\|\mathbf{x}\|_2^2
  \end{equation}
for all $K$-sparse vectors $\mathbf{x}$. Specifically, the minimum of all constants $\delta$'s satisfying \eqref{rip} is called the isometry constant and denoted as $\delta_K$.
 \end{definition}
 
The RIP characterizes how close the matrix $\mathbf{\Phi}$ behalves like an identity matrix, which has been popularly used for the analyses of  CS reconstruction algorithms~\cite{candes2005decoding,candes2008restricted}.

 \begin{definition}(Tight Frame~\cite{Strohmer2003grassmannianframes}) \label{def:tf}
    A family $\mathbf{F} = \{ \mathbf{f}_i\}_{i \in \Omega} $ in a Hilbert space ${\mathbb{H}}$ is called a frame if there exist two constants $0 < \alpha \leq \beta < \infty$ obeying the following condition,
    \begin{equation}
    \begin{array}{*{20}{c}}
     \alpha \| \mathbf{ f } \|_2^2 \leq \sum \limits_{ i \in \Omega}  | \langle \mathbf{f}, \mathbf{f}_i \rangle |^2 \leq \beta \| \mathbf{f} \|_2^2,  &  \forall \mathbf{f} \in \mathbb{R}^m,
    \end{array}
    \end{equation}
    where $\alpha$ and $\beta$ are called frame bounds. $\alpha$-tight frame is a frame whose bounds are equal, namely $\alpha = \beta$.
    Specially, a tight frame with $\alpha = 1$ is called a Parseval tight frame.
    \end{definition}

\subsection{Lemmas}
 

 The following lemmas are useful for our analysis.

\begin{lemma}(Singular values of random Gaussian matrices) \label{sigma of Gauss matrix}
Given $m,n \in \mathbb{N}$ with $m \leq n$ and let $\mathbf{\Psi} \in {\mathbb{R}^{m \times n}}$ be a Gaussian random matrix with entries ${\varphi _{i,j}}\mathop  \sim \limits^{i.i.d}  \mathcal{N} ( 0, {1}/{m} )$. Let $\sigma_1(\mathbf{\Psi}) \ge \cdots \ge \sigma_m (\mathbf{\Psi})$ be the singular values of $\mathbf{\Psi}$. Then, for any constant $\varepsilon \geq 0$,
\begin{subequations}
\begin{align}
       \Pr \big\{ \sigma_1 (\mathbf{\Psi} )  \geq \sqrt{n/m} + 1+ \varepsilon  \big \} & \leq e^ {- m \varepsilon^2 / 2},\\
       \Pr \big \{ \sigma_m (\mathbf{\Psi}) \leq  \sqrt{n/m} -  1  - \varepsilon \big \} & \leq e^ {- m \varepsilon^2 / 2}.  
\end{align} 
\end{subequations}
\end{lemma}

It is worth mentioning that the concentration inequalities in Lemma~\ref{sigma of Gauss matrix} can be easily derived from~\cite[Section~3]{candes2005decoding}, as they differs only in the variance of Gaussian entries. Specifically, the entries of $\mathbf{\Psi}$  in Lemma~\ref{sigma of Gauss matrix} are drawn {\it i.i.d.} from $\mathcal{N} ( 0, {1}/{m} )$, while those in~\cite[Section 3]{candes2005decoding} are {\it i.i.d.} from $\mathcal{N} ( 0, {1}/{n} )$. See Appendix~\ref{app:0} for more details. 


\begin{lemma}(Tail bound for random Gaussian matrix~\cite{Chen2016influencesofpreconditioningonmutualcoherence}) \label{mutual coherence of Gauss matrix}
Given $m,n \in \mathbb{N}$ with $m \leq n$ and let $\mathbf{\Psi} \in {\mathbb{R}^{m \times n}}$ be a random Gaussian matrix with entries ${\varphi _{i,j}}\mathop  \sim \limits^{i.i.d} \left( {0,{1}/{m}} \right)$. Then for any constant $\eta \in (0,1)$, the mutual coherence of $\mathbf{\Psi}$ satisfies
\begin{eqnarray}
~ \Pr \left \{ \mu ( \mathbf{\Psi} )  \geq   \eta \right \}   \leq   n (n-1)  \big( e^{ - m \eta^2 / ( 16 + 4 \eta )}   +  e^{ - m / 16 }  \big). \hspace{-1mm}
\end{eqnarray}
\end{lemma}


\begin{lemma}(Generalized Wielandt inequality~\cite{lin2012generalizedWielandtinequality}) \label{wielandt inequality}
Let $\mathbf{A}\in \mathbb{R}^{n\times n}$ be a  positive-definite matrix with eigenvalues $\lambda_1 \geq \cdots \geq \lambda_n$ and eigenvectors $\mathbf{x}_1, \cdots, \mathbf{x}_n$ of unit length. Let $\mathbf{u}, \mathbf{v} \in \mathbb{R}^n$ be nonzero vectors and $\varphi:  = \frac{ |\langle\mathbf{u}, \mathbf{v} \rangle| }{{\left\| \mathbf{u} \right\|_2 \left\| \mathbf{v} \right\|_2}}.$ Then,  
\begin{equation}
\left| {{\mathbf{u}^{T}}\mathbf{A} \mathbf{v}} \right|^2 \hspace{-.5mm} \leq \hspace{-.5mm} {\left( {\frac{{{\lambda _1} - {\lambda _n} + \left( {{\lambda _1} + {\lambda _n}} \right) \varphi }}{{{\lambda _1} + {\lambda _n} + \left( {{\lambda _1} - {\lambda _n}} \right) \varphi }}} \right)^2}{\mathbf{u}^{T}}\mathbf{A} \mathbf{u} {\mathbf{v}^{T}}\mathbf{A} \mathbf{v}. \label{ineqality: wielandt}
\end{equation} 
In particular, the equality 
holds if and only if either the vectors $\mathbf{u}$ and $\mathbf{v}$ are collinear or of the following form:
\begin{equation}
\begin{cases}
 \mathbf{u}  = {\left\| \mathbf{u} \right\|_2}\left( {s_1 \sqrt {1 + \varphi } {\mathbf{x}_1} + s_2 \sqrt {1 - \varphi  } {\mathbf{x}_n}} \right) / {\sqrt 2 }, \\ 
 \mathbf{v} = {\left\| \mathbf{v} \right\|_2} \left( {s_1 \sqrt {1 + \varphi } {\mathbf{x}_1} - s_2 \sqrt {1 - \varphi } {\mathbf{x}_n}} \right) s_3 /{\sqrt 2 },  
\end{cases}
\end{equation}
where $s_1$, $s_2$ and $s_3$ take values from $\{1, -1\}$.
\end{lemma}

 
 \begin{lemma} (Consequences of the RIP \cite{Dai2009subspacepursuit,kwon2014MMP}])  \label{rip:consequence} Let $\mathcal{S}\subseteq\Omega$ and let $\mathbf{\Phi}$ be a matrix satisfying the RIP with isometry constant $\delta_{|\mathcal{S}|}<1$. Then for any vector $\mathbf{u}\in \mathbb{R}^{|\mathcal{S}|}$,
 \begin{subequations}
\begin{align}
&~~~~~~~~~~~~~~~~~~~~~~\delta_{|\mathcal{S}|} \le (|\mathcal{S}|-1)\mu, \label{rip:consequence:3}\\
      &~~~~~~~~~~~\frac{\|\mathbf{u}\|_2}{1+ \delta_{|\mathcal{S}|}}\le\|(\mathbf{\Phi}_{\mathcal{S}}^{T}\mathbf{\Phi}_{\mathcal{S}})^{-1} \mathbf{u}\|_2\le\frac{\|\mathbf{u}\|_2}{1 - \delta_{|\mathcal{S}|}}, \label{rip:consequence:2}\\
      &~~~~~(1-\delta_{| \mathcal{S} |})\|\mathbf{u}\|_2\le\|\mathbf{\Phi}_{\mathcal{S}}^{T}\mathbf{\Phi}_{\mathcal{S}}  \mathbf{u}\|_2\le(1 + \delta_{| \mathcal{S} |})\|\mathbf{u}\|_2, \label{rip:consequence:1}\\
& \hspace{-2mm}  (1 \hspace{-.25mm}  -  \hspace{-.25mm} \delta_{| \mathcal{S} |})   \big\|  \big(  \mathbf{\Phi}_{\mathcal{S}}^\dag   \big) ^{T} \hspace{-.25mm} \mathbf{u} \big\|_2^2
\hspace{-.25mm}  \leq \hspace{-.25mm} \| \mathbf{u} \|_2^2 \hspace{-.25mm}  \leq  \hspace{-.25mm}  (1 \hspace{-.25mm} +  \hspace{-.25mm} \delta_{| \mathcal{S} |})   \big\|  \big(  \mathbf{\Phi}_{\mathcal{S}}^\dag   \big) ^{T} \hspace{-.25mm}  \mathbf{u} \big\|_2^2. \hspace{-1mm}
 \label{rip:consequence:4} 
 \end{align}
\end{subequations}
 \end{lemma}


\begin{lemma}(Norm inequality~\cite{golub1996matrixcomputations}) \label{norminquality} For matrices $\mathbf{A}, \mathbf{B} \in \mathbb{R}^{m\times n}$ and $\mathbf{u} \in \mathbb{R}^{m}$, it is satisfied that
 \begin{subequations}
\begin{flalign}
   &    \big| \|\mathbf{A}\|_2-\|\mathbf{B}\|_2 \big| \leq \|\mathbf{A}+\mathbf{B}\|_2 \leq \|\mathbf{A}\|_2+\|\mathbf{B}\|_2,  \label{norminquality:1} \\ 
   & \hspace{18.75mm}  \frac{\|\mathbf{u}\|_1}{\sqrt{m}}\le \|\mathbf{u}\|_2 \le \sqrt{m}\|\mathbf{u}\|_{\max},  \label{norminquality:2}  
\end{flalign}
\end{subequations}
where $\|\mathbf{u}\|_{\max} = \max_{i} |u_i|$. 
\end{lemma}


  \begin{lemma} 
   \label{lem:mip_1}
For two disjoint sets $I_1, I_2 \subset \Omega$ and $\mathbf{\Phi}_{I_1}, \mathbf{\Phi}_{I_2}$, which are two subsets of matrix $\mathbf{\Phi} \in \mathbb{R}^{m \times n}$,  it holds that
\begin{equation}
    \|\mathbf{\Phi}_{I_1}^{T}\mathbf{\Phi}_{I_2}\|_2 \leq  \sqrt{|I_1| \hspace{-.5mm} \cdot \hspace{-.5mm} |I_2|} \mu (\mathbf{\Phi})  .
\end{equation}
\end{lemma}

\begin{IEEEproof} We have
\begin{eqnarray} \label{eq:lem_disj}
    \|\mathbf{\Phi}_{I_1}^{T}\mathbf{\Phi}_{I_2}\|_2 & \leq & \sqrt{ | I_1 | | I_2 | } \|\mathbf{\Phi}_{I_1}^{T}\mathbf{\Phi}_{I_2}\|_{\max}   \nonumber   \\
        &=& \sqrt{ | I_1 | | I_2 | } \hspace{.25mm} \mu \left( [\mathbf{\Phi}_{I_1} \mathbf{\Phi}_{I_2} ] \right)  \nonumber   \\
        &\leq& \sqrt{ | I_1 | | I_2 | } \hspace{.25mm} \mu (\mathbf{\Phi}),
\end{eqnarray} 
which is the desired result.
\end{IEEEproof}

\begin{lemma}(\hspace{-.001mm}\cite[Lemma 3]{wen2017OLS}) \label{projection} 
Suppose that $\mathcal{S}\subset \Omega$ and let $\mathbf{\Phi}$ have unit $l_2$-norm columns and satisfy the $RIP$ of order $|\mathcal{S}| + 1$. Then, for any $i \in \{1,2,\cdots,n\} \backslash \mathcal{S}$,
\begin{equation} \label{lemma:OlS}
\|\mathcal{P}_{\mathcal{S}}^\perp \mathbf{\phi}_i\|_2 \geq \sqrt{1-\delta^2_{|\mathcal{S}|+1}}.
\end{equation}
\end{lemma}
The lower bound for $\|\mathcal{P}_{\mathcal{S}}^\perp \mathbf{\phi}_i\|_2$  has also been shown to be sharp in~\cite{wen2017OLS}.

\section{The PmOLS Algorithm}  \label{sec:III}

In this section, we theoretically study the performance of PmOLS. As shown in Algorithm~\ref{tab:pmols},  the PmOLS algorithm consists of two main parts: i) preconditioning and ii)  signal reconstruction. Specifically, the first part uses  the PIP method to reduce the mutual coherence of sampling matrix, while the second part employs the conventional mOLS algorithm~\cite{wang2017MOLS} to reconstruct sparse signals. 

\subsection{The PIP Method}

The key idea of the PIP method is to multiply a preconditioner $\mathbf{P} \in \mathbb{R} ^{n \times m}$ to the sampling matrix $\mathbf{\Psi}$ so as to make the matrix $\mathbf{P} \mathbf{\Psi}$ approach an identity matrix:
\begin{equation} \label{min:P}
\mathbf{P} = \mathop {\arg \min} \limits_{\mathbf{P}} {\left\| \mathbf{P} \mathbf{\Psi} - \mathbf{I} \right\|_F},
\end{equation}
where $\|\cdot \|_F$ denotes the Frobenius norm and $\mathbf{I} \in \mathbb{R}^{n \times n}$ is an identity matrix. The solution of problem~\eqref{min:P} is given in the following proposition. 

\begin{proposition} \label{prop:P}
Let  $\mathbf{U}$ and $\mathbf{V}$ be the left- and right-singular matrices of $\mathbf{\Psi} \in \mathbb{R}^{m \times n}$, respectively, and let $\sigma_1, \cdots, \sigma_r$ be the nonzero singular values of $\mathbf{\Psi}$ in descending order. Then, the solution to~\eqref{min:P} can be given by
\begin{equation}
\mathbf{P} \hspace{-.25mm}  = \hspace{-.5mm}
\begin{cases}
 \mathbf{\Psi}^T \left( \mathbf{\Psi} \mathbf{\Psi}^T \right) ^{-1}                &  \hspace{-1.25mm} r = m, \\
\left( \mathbf{\Psi}^T \mathbf{\Psi} \right) ^{-1} \mathbf{\Psi}^T.                &  \hspace{-1.25mm} r = n, \\
\mathbf{V}   \text{diag} \big( \hspace{-.25mm} \big[ \hspace{-.5mm} \underbrace{ {1}/{\sigma_1}, \hspace{-.25mm}\cdots \hspace{-.5mm}, {1}/{\sigma_r}}_{r},\underbrace{0, \hspace{-.5mm} \cdots \hspace{-.5mm}, 0}_{n - r} \big] \hspace{-.25mm} \big) \hspace{-.25mm} \mathbf{U}^T  & \hspace{-1.25mm} r < \min \{m, n\}. \vspace{-3mm}
\end{cases}  
\end{equation}      
Moreover, 
\begin{equation} \label{eq:itit}
    \mathbf{P} \mathbf{\Psi} = \mathbf{V}   \text{diag} \big( [\underbrace{1,\cdots, 1}_{r},\underbrace{0,\cdots,0}_{n - r} ] \big) \mathbf{V}^T.
\end{equation}

\end{proposition}
\begin{IEEEproof}
See Appendix \ref{app:P}.
\end{IEEEproof}

This proposition provides closed-form solutions to~\eqref{min:P} pertaining to three cases. 
To analyze the PIP method, however, we are interested in the case of $m < n$, which is a typical down-sampling of GISC. The following theorem characterizes the mutual coherence of the preconditioned sampling matrix via PIP.

\begin{theorem}\label{theorem: mu of preconditioned matrix}
Let $\mathbf{\Psi}\in \mathbb{R}^{m\times n}$ be a random Gaussian matrix with $m < n$ and entries ${\varphi _{ij}}\mathop  \sim \limits^{i.i.d} \mathcal{N}  (0,{1}/{m} )$ and $ \mathbf{P} := \mathbf{\Psi}^{T} ( \mathbf{\Psi} \mathbf{\Psi}^{T} ) ^{-1}$ be the preconditioner of the PIP method. Also, let $\mu(\mathbf{P} \mathbf{\Psi})$ be the mutual coherence of matrix $\mathbf{P} \mathbf{\Psi}$. Then, for any constant $\eta \in (0, 1)$, we have
\begin{equation}
\Pr (\mu(\mathbf{P} \mathbf{\Psi}) \leq \eta) \geq 1 - 3 n^2 e^{ -  {m \eta^2}/{ 72 } }.
 \label{probability:mu}
\end{equation} 
\end{theorem}
\begin{IEEEproof}
See Appendix \ref{app:1}.
\end{IEEEproof} 

\vspace{1mm}
One can interpret from Theorem \ref{theorem: mu of preconditioned matrix} that the probability in the right-hand side of~\eqref{probability:mu} increases monotonically with the problem dimension. Recently, it has brought to our attention that under the down-sampling setting, generalized alternating projection (GAP)~\cite{Liao2014generalized,Yuan2018adaptive} implied a similar strategy as the one posted in Proposition~\ref{prop:P}. Nevertheless, our method  is more general as it covers all possible ranks of sampling matrices. Moreover, we establish a probabilistic guarantee for our method, for which there is no counterpart in the GAP studies.

\subsection{The mOLS Algorithm}  
For notational simplicity, denote  $ \mathbf{y} := \mathbf{P}\mathbf{y}_0$ and $\mathbf{\Phi} :=  \mathbf{P}\mathbf{\Psi}$, which we call the preconditioned samples and the preconditioned sampling matrix, respectively. Then, the system model can be rewritten as 
                 \begin{equation}
                 \mathbf{y} = \mathbf{\Phi} \mathbf{x}. 
                 \end{equation}
                 
 In the signal reconstruction step, the mOLS algorithm solves problem~\eqref{equation: preconditin_measurement} in an iterative manner. Specifically, in the $(k + 1)$-th iteration ($k \geq 0$), it adds $s$ indices to the previously selected index set $\mathcal{S}^k$ to obtain
\begin{equation} \label{mols:iden}
  \mathcal{S}^{k+1} = \mathcal{S}^{k} \cup \mathop {\arg \min }\limits_{\mathcal{S}: |\mathcal{S}|= s}\sum\limits_{i \in {\mathcal{S}}} \big \| \mathcal{P}_{\mathcal{S}^k \cup{\{i\}}}^\bot \mathbf{r}^k \big \|_2.
\end{equation}
The vestige lying in the subspace spanned by the columns of $\mathbf{\Phi}_{\mathcal{S}^{k + 1}}$ is then eliminated from $\mathbf{y}$, yielding the residual vector $\mathbf{r}^{k + 1}$ for the $(k + 1)$-th iteration.  Put formally, 
\begin{eqnarray}
\mathbf{r}^{k + 1}& \hspace{-1mm} = & \hspace{-1mm} \mathbf{y} - \mathbf{\Phi} \mathbf{x}^{k + 1} \nonumber \\
& \hspace{-1mm} = & \hspace{-1mm} \mathbf{y} - \mathbf{\Phi}_{ \mathcal{S}^{k + 1} } \mathbf{\Phi}_{ \mathcal{S}^{k + 1}} ^ \dag \mathbf{y} \nonumber \\
& \hspace{-1mm} = & \hspace{-1mm} \mathbf{y} - \mathcal{P}_{ \mathcal{S}^{k + 1} } \mathbf{y} =   \mathcal{P}_{\mathcal{S}^{k+1}}^\bot \mathbf{y},  \label{eq:rrrrr}
\end{eqnarray} 
where the second equality is because 
\begin{equation}
 (\mathbf{x}^{k + 1})_{ \mathcal{S}^{k + 1} }  =  \mathbf{\Phi}_{ \mathcal{S}^{k + 1} } ^ \dag \mathbf{y} ~~\text{and}~~ (\mathbf{x}^{k + 1})_{ \Omega \backslash \mathcal{S}^{k + 1} }  =  \mathbf{0}.
\end{equation}
The expression of the residual $\mathbf{r}^{k + 1}$ will play an important role in our  analysis in Appendix~\ref{app:2}. 

%

As mentioned in~\cite{wang2017MOLS}, the straightforward implementation of~\eqref{mols:iden} is computationally prohibited as it computes and sorts the  elements in $\{  \| \mathcal{P}_{\mathcal{S}^k \cup \{ i\} }^ \bot \mathbf{y} \| _2  \}_{i \in \Omega \backslash \mathcal{S}^k}$ to find out the smallest $s$ ones, which involves $n - sk$ orthogonal projections (i.e., $\mathcal{P}_{\mathcal{S}^k \cup{\{i\}}}^\perp, \forall i \in \Omega \backslash \mathcal{S}^k$). Nevertheless, following~\cite{Blumensath2007differentols}, a more efficient way for solving~\eqref{mols:iden} is
\begin{eqnarray}
  \mathcal{S}^{k+1} = \mathcal{S}^{k} \cup \mathop {\arg \max }\limits_{\mathcal{S}: \left| \mathcal{S} \right| = s} \sum\limits_{i \in \mathcal{S}} {\frac{{\left| {\left\langle {{\phi _i},{\mathbf{r}^k}} \right\rangle } \right|}}{{{{\hspace{1.75mm}\left\| {\mathcal{P}_{{\mathcal{S}^k}}^ \bot {\phi _i}} \right\|}_2}}}},  \label{mols:re1} 
\end{eqnarray}
which is much simpler than \eqref{mols:iden} as it requires only the projection operator  $\mathcal{P}_{{\mathcal{S}^k}}^ \bot$. 

Moreover, from the property of orthogonal complement projector (i.e., $\mathcal{P}_{{\mathcal{S}^k}}^ \bot = ( \mathcal{P}_{{\mathcal{S}^k}}^ \bot )^T = ( \mathcal{P}_{{\mathcal{S}^k}}^ \bot )^2$),  the correlation term $\left| {\left\langle {{\phi _i},{\mathbf{r}^k}} \right\rangle } \right|$ in the right-hand side of \eqref{mols:re1} satisfies
\begin{eqnarray}
\left| {\left\langle {{\phi _i},{\mathbf{r}^k}} \right\rangle } \right|  & \hspace{-2mm }\overset{\eqref{eq:rrrrr}}{=} & \hspace{-2mm}\left| {\left\langle {{\phi _i}, \mathcal{P}_{\mathcal{S}^{k}}^\bot \mathbf{y} } \right\rangle } \right| \nonumber \\
&\hspace{-2mm}  = & \hspace{-2mm} \left| {\left\langle {\phi _i}, ( \mathcal{P}_{\mathcal{S}^{k}}^\bot )^2 \mathbf{y}   \right \rangle } \right| \nonumber \\
&\hspace{-2mm}  = & \hspace{-2mm} \left| { \left\langle  {\phi _i},  (\mathcal{P}_{\mathcal{S}^{k}}^\bot )^T  \mathcal{P}_{\mathcal{S}^{k}}^\bot  \mathbf{y}   \right \rangle } \right|  \nonumber \\
&\hspace{-2mm}  = & \hspace{-2mm} \left| { \left\langle  \mathcal{P}_{\mathcal{S}^{k}}^\bot {\phi _i},  \mathcal{P}_{\mathcal{S}^{k}}^\bot  \mathbf{y}   \right \rangle } \right| \nonumber \\
&\hspace{-2mm}  = & \hspace{-2mm} \left| { \left\langle  \mathcal{P}_{\mathcal{S}^{k}}^\bot {\phi _i},  \mathbf{r}^k   \right \rangle } \right|. 
\end{eqnarray}
Thus, one can rewritten \eqref{mols:re1} as 
\begin{equation}
  \mathcal{S}^{k+1} = \mathcal{S}^{k} \cup  \underset{\mathcal{S} : | \mathcal{S} | =L} {\arg \max}  \sum_{i \in \mathcal{S}}
\bigg| \bigg\langle \frac{ \mathcal{P}^{\bot}_{\mathcal{S}^{k}}
\phi_{i} }{\| \mathcal{P}^{\bot}_{\mathcal{S}^{k}} \phi_{i} \|_{2}},
\mathbf{r}^k \bigg\rangle \bigg|, 
\end{equation}
which implies that the behavior of mOLS is un-changed whether the sampling matrix is normalized or not. 

The following theorem gives a recovery condition for the mOLS algorithm to perfectly recover any $K$-sparse signal. 


\begin{theorem} \label{theorem: mols}
Consider the mOLS algorithm with selection parameter $s$. Let $\mathbf{\Phi} \in {\mathbb{R} ^{m \times n }}$ be a sampling matrix with unit $\ell_2$-norm columns. Then,
 mOLS exactly recovers any vector $\mathbf{x}$  from its samples $\mathbf{y} = \mathbf{\Phi} \mathbf{x}$ if  the mutual coherence of matrix $\mathbf{\Phi}$ satisfies
\begin{equation} \label{bounds:mols}
\mu ({\bf \Phi}) < \frac{1} {2sK - 2s + 1}.
\end{equation} 
\label{theorem: mols}
\end{theorem}
\begin{IEEEproof}
See Appendix \ref{app:2}.
\end{IEEEproof}
\vspace{1mm}
The mOLS algorithm reduces to the conventional OLS algorithm when $s = 1$ and the recovery condition becomes 
\begin{equation} \label{bounds:ols}
\mu ({\bf \Phi}) < \frac{1} {2K - 1},
\end{equation}
which matches the recovery condition of OLS proposed in~\cite{Soussen2013OLSsufficient}. One can notice from~\eqref{bounds:mols} that the recovery condition becomes more restrictive as the parameter $s$ increases. This in turn implies that the sparsity range of signals that is guaranteed to be recovered via mOLS narrows down as $s$ goes large. 
On the other hand, a larger $s$ may accelerate the convergence rate of the algorithm. Therefore, an appropriate  choice of $s$ to trade off the convergence rate and recovery guarantee of the mOLS algorithm is of vital importance; See more discussions in~\cite{wang2017MOLS}. 

%

\subsection{Exact  Recovery via PmOLS}

Thus far, we have discussed the performance guarantees of PIP and mOLS in Theorem~\ref{theorem: mu of preconditioned matrix} and  Theorem~\ref{theorem: mols}, respectively. By relating these two theorems, we obtain the recovery condition for the PmOLS algorithm.

\begin{theorem} \label{theorem:pmols}
Consider the PmOLS algorithm with selection parameter $s$. Let $\mathbf{\Psi}\in \mathbb{R}^{m\times n}$ be a random Gaussian sampling matrix with $m < n$ and entries ${\varphi _{ij}}\mathop  \sim \limits^{i.i.d} \mathcal{N} ( {0,{1}/{m}} )$ and $ \mathbf{P} := \mathbf{\Psi}^{T} ( \mathbf{\Psi} \mathbf{\Psi}^{T} ) ^{-1}$ be the preconditioner of the PIP method. 
Then, PmOLS perfectly recovers any $K$-sparse vector $\mathbf{x}$ from the preconditioned samples $\mathbf{y} = \mathbf{P \Psi} \mathbf{x}$ with probability exceeding  
\begin{eqnarray}
 1 - 3 n^2   e^{  - m / [ 72 ( 2 K s - 2 s + 1 ) ^2 ] }. \label{eq:25}
\end{eqnarray}

\end{theorem}
\begin{IEEEproof}
See Appendix \ref{app:3}.
\end{IEEEproof}

Clearly Theorem~\ref{theorem:pmols} implies that the choice of 
\begin{equation} \label{eq:mmm}
m \geq c K^2 \log (n/\epsilon)
\end{equation}
 is sufficient to ensure the success probability of PmOLS exceeding $1 - \epsilon$. We would like to point out that the suggested sampling complexity in~\eqref{eq:mmm} cannot be fundamentally improved. Specifically, it has been shown in~\cite{wen2017OLS} that OLS (i.e., the mOLS algorithm with  $s = 1$) may fail to recover some $K$-sparse signal under 
\begin{equation}
\delta_{K + 1} = \frac{1}{\sqrt {K + \frac{1}{4}}}.
\end{equation}
For $m \times n$ random Gaussian matrices with entries drawn
{\it i.i.d.} from Gaussian distribution $\mathcal{N}(0, {1}/{m})$, if one wishes to ensure the above RIP condition with high probability, then it requires at least~\cite{candes2005decoding}
\begin{equation}
 m \geq cK^2 \log (n / K),
 \end{equation} which remains a gap to the result in \eqref{eq:mmm}. Whether it is possible to bridge this gap is an interesting open question. 

The suggested bound in \eqref{eq:mmm} is closely related to~\cite{tropp2007signal}, where it is shown that OMP can exactly recover a given $K$-sparse $n$-dimensional from its 
\begin{equation} \label{eq:mmm2}
m \geq c K \log (n/\epsilon)
\end{equation}
 random Gaussian samples with probability $1 - \epsilon$. The difference in the sampling complexity is due to that~\eqref{eq:mmm2} is derived for the recovery of a given sparse signal (i.e., a sparse signal with the fixed support); Whereas, our result in \eqref{eq:mmm} holds uniformly for all $K$-sparse signals.

%

\section{Simulation and Experimental Results} \label{sec:V}

\subsection{Simulation Results}

In this section, in order to evaluate the empirical performance of the PmOLS algorithm,  we carry out numerical simulations.  
The first simulation aims to show how effective the PIP method is in reducing the mutual coherence of sampling matrices. To the end, we perform $500$ independent trials to compare the mutual coherence of sampling matrices before and after preconditioning. 

\begin{figure*}[!htb]
\centering
\subfigure[Comparison of mutual coherence]
{\hspace{-6mm} \includegraphics[width = 85mm, height = 85mm] {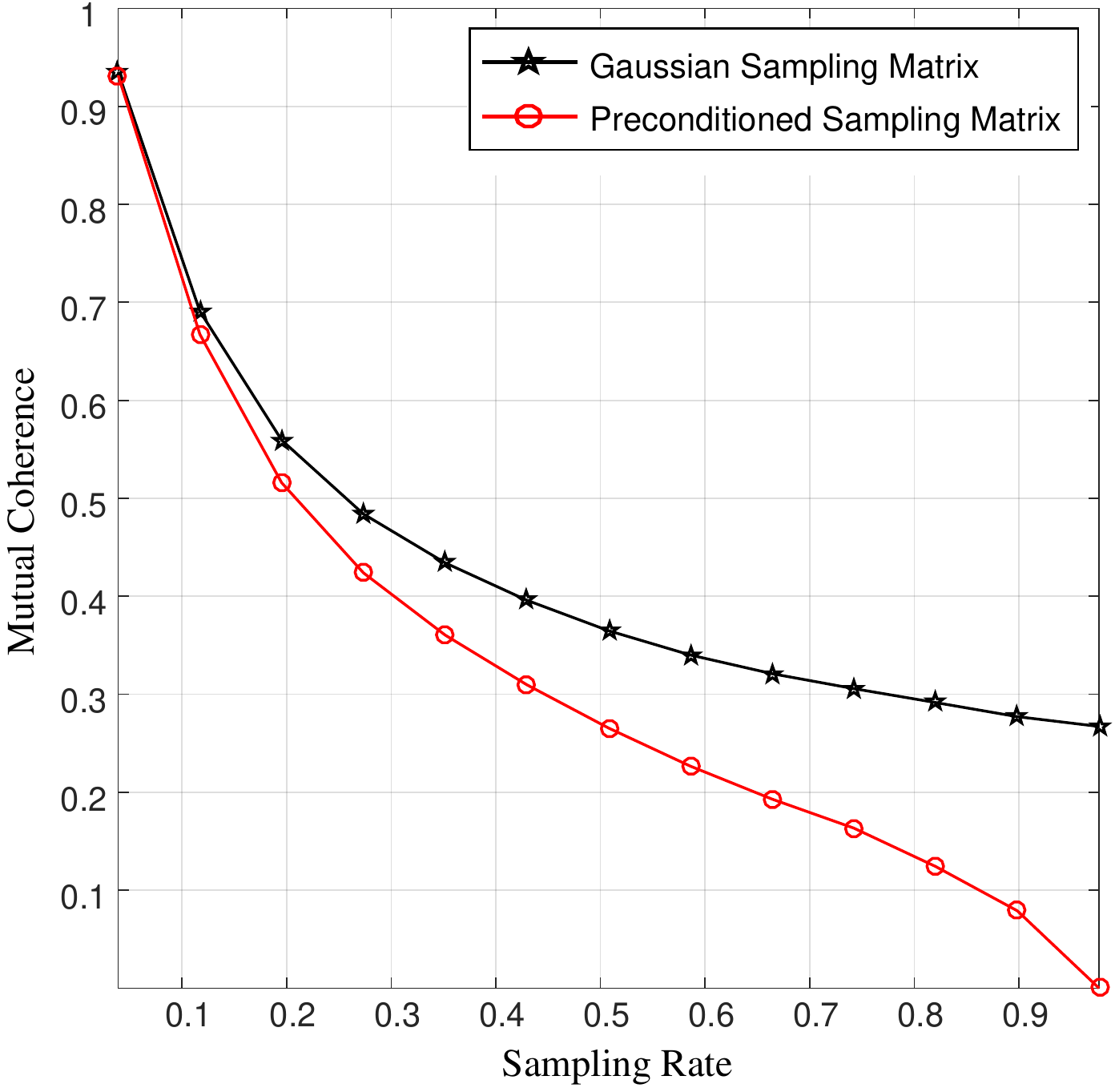}
 \label{fig:mutual coherence} \hspace{5mm} }
\subfigure[Frequency of exact recovery of sparse Gaussian signals]
{\includegraphics[width =85mm, height = 85mm] {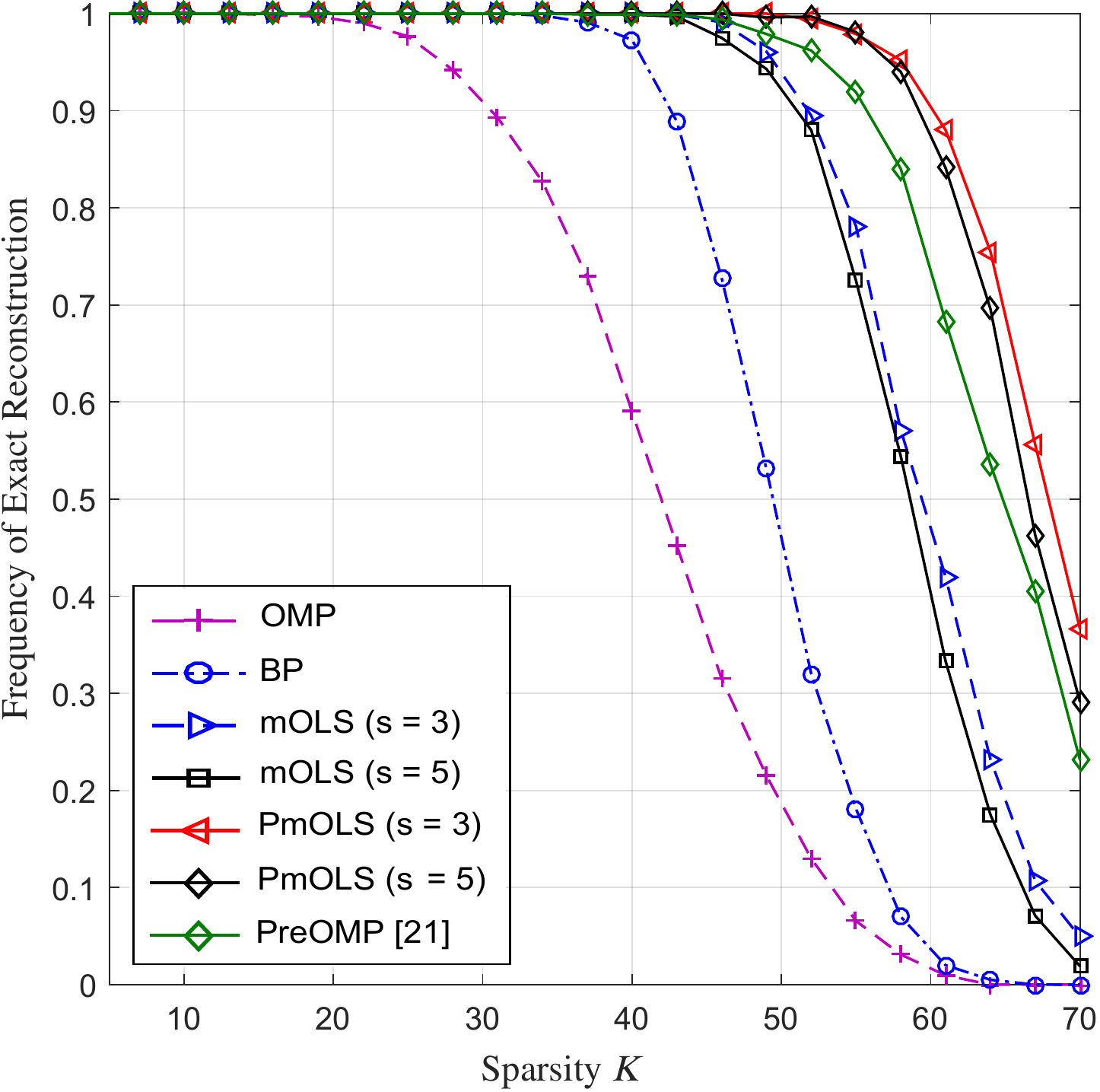} 
\label{fig1:Gauss} }
\subfigure[Frequency of exact recovery of sparse $2$-PAM signals]
{\hspace{-6mm} \includegraphics[width =85mm, height = 85mm] {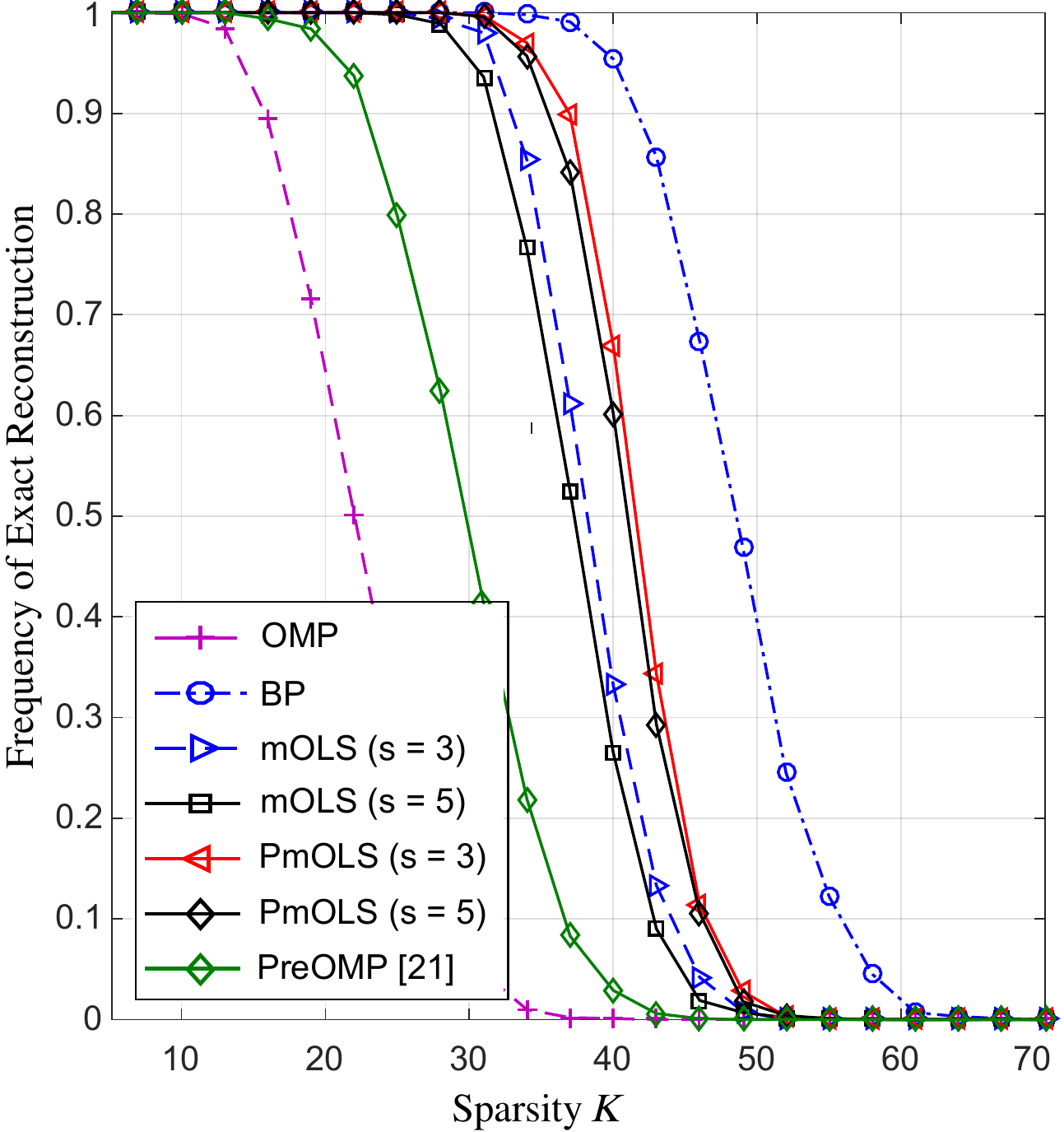} 
\label{fig1:PAM} \hspace{6.5mm}}
\subfigure[Frequency of exact recovery of sparse two-valued signals] 
{\includegraphics[width =85mm, height = 85mm] {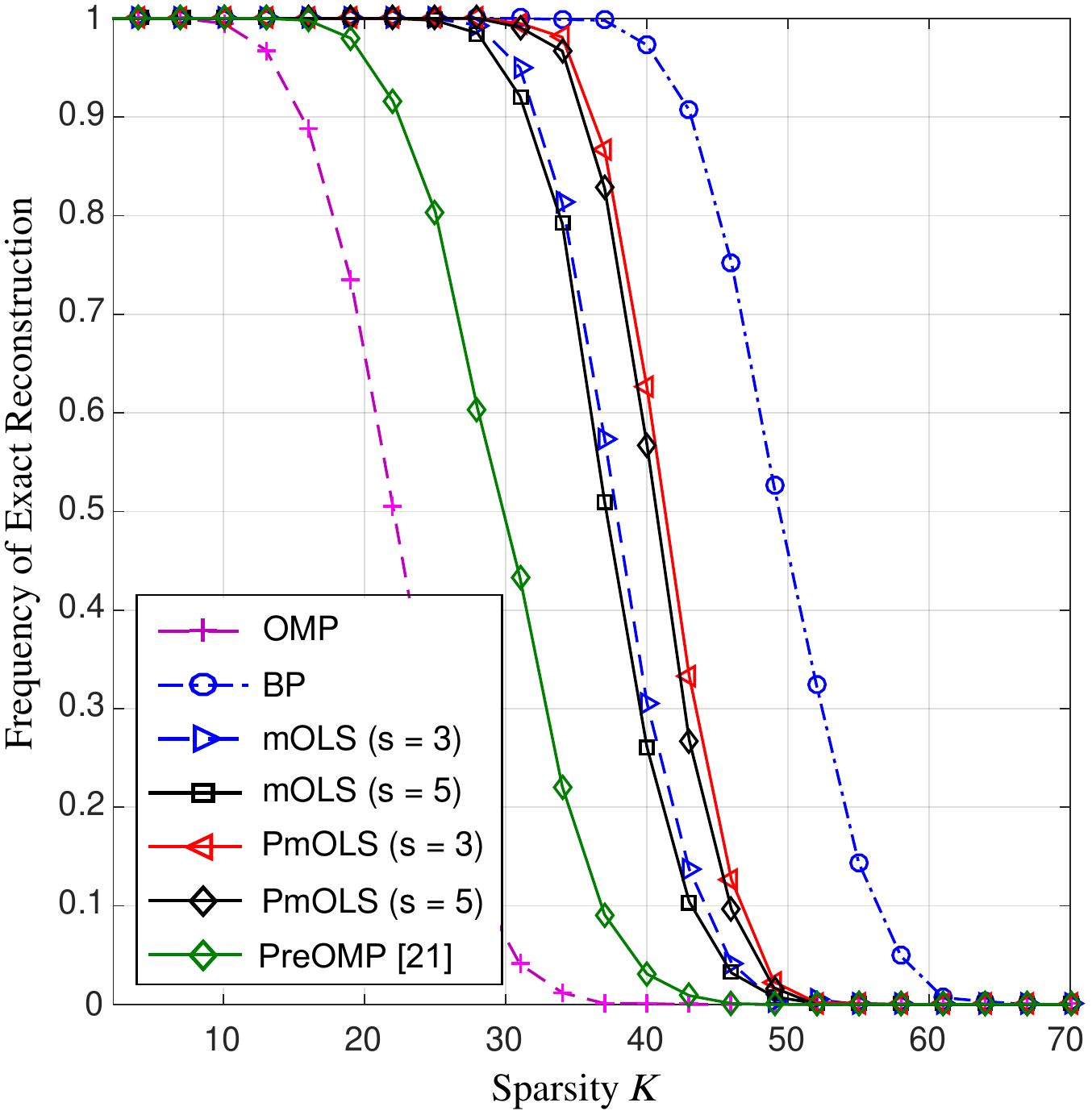} 
\label{fig1:POS}} 
\caption{Simulation results: (a) mutual coherence as a function of sampling rate;  (b)--(d) Frequency of exact recovery of three types of sparse signals as a function of $K$ for the random Gaussian sampling matrix.} 
\label{fig:ER1}   
\end{figure*}

In each trial, we construct an $m \times n$ random Gaussian matrix $\mathbf{\Psi}$ with entries drawn {\it i.i.d.} from Gaussian distribution ${\mathcal N}(0,{1}/{m})$, where we fix $n = 256$ and let the sampling rate vary from $5\%$ to one. For each sampling rate ${m}/{n}$, we compute the mutual coherence of the original sampling matrix $\mathbf{\Psi}$ and that of the preconditioned sampling matrix $\mathbf{P} \mathbf{\Psi}$, respectively. By averaging the values over $500$ independent trials, we plot the results as a function of sampling rate in Fig.~\ref{fig:mutual coherence}. It can be observed that for both matrices $\mathbf{\Psi}$ and $\mathbf{P} \mathbf{\Psi}$,  the mutual coherence decreases as the sampling rate increases. In particular, the mutual coherence of $\mathbf{ P } \mathbf{ \Psi }$ is uniformly better than that of $\mathbf{ \Psi }$ and the gap between them increases with the sampling rate, which clearly demonstrates the effectiveness of the proposed PIP method.  

\begin{figure}[t]
\centering \hspace{4mm}{\includegraphics[width =85 mm]{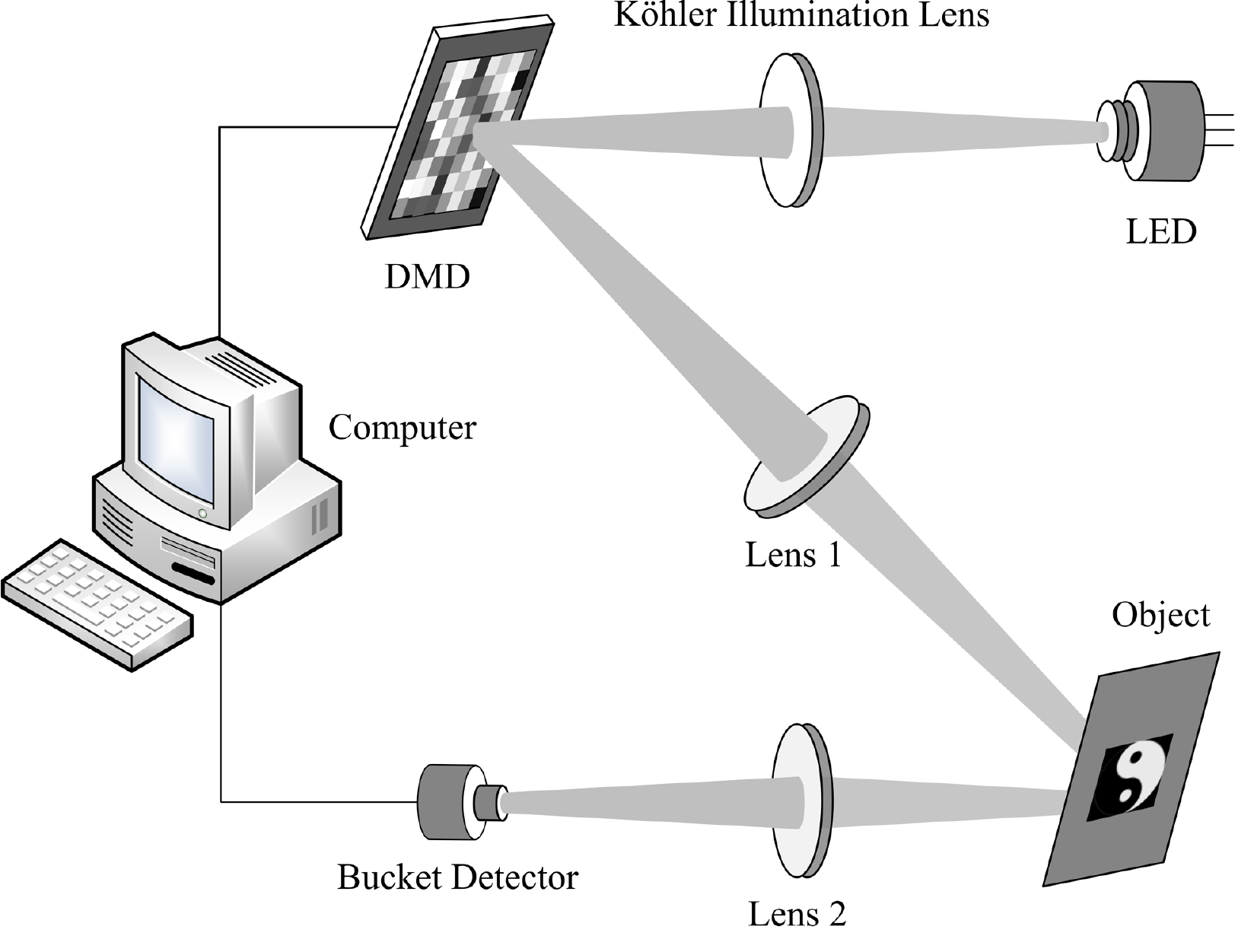}}
\caption{System diagram of GISC based on the digital micromirror device (DMD).} \label{fig:setup}    \vspace{-3mm}
\end{figure}

\begin{figure*}[t]
\centering 
{\includegraphics[width = 150mm] {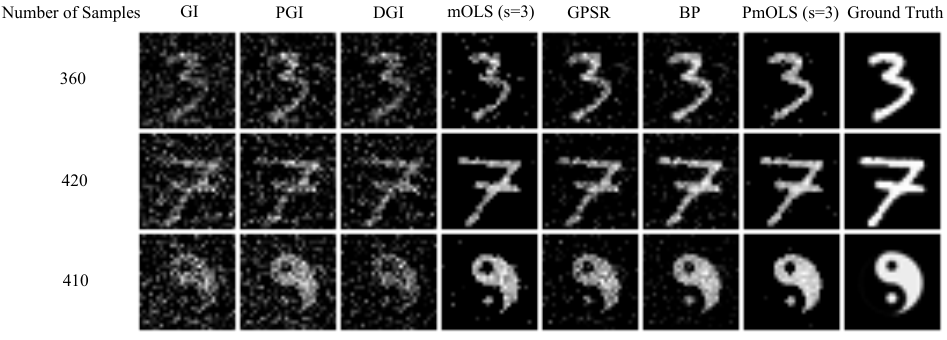}}
\label{fig:Gaussreconstruction}
\caption{Experimental results: the number of samples under test and the ground truth objects are shown in the leftmost and rightmost column, respectively. The $2$--$7$ columns in the middle are the reconstruction results for the GI, PGI, DGI, mOLS, BP and PmOLS approaches, where different objects are imaged and compared in different rows.} 
\label{fig:experimental results} 
\end{figure*}

In the second part of our simulations, we empirically compare the recovery performance of the PmOLS algorithm with other approaches. We follow the simulation strategy in~\cite{Dai2009subspacepursuit}, which evaluates the effectiveness of recovery algorithms by checking the empirical frequency of exact reconstruction of sparse signals. In particular,  by comparing the maximum sparsity level, i.e., the critical sparsity~\cite{Dai2009subspacepursuit} at which exact reconstruction of sparse signals is always guaranteed, the recovery performance of different approaches can be empirically compared.
We consider  random Gaussian matrices of size $128 \times 256$ with entries drawn {\it i.i.d.} from Gaussian distribution ${\mathcal N}(0, {1}/{128})$. Also, we consider three types of signals to be recovered: i)  sparse Gaussian signals, ii) sparse pulse amplitude modulation (PAM) signals and iii)  sparse two-valued signals, whose nonzero elements are drawn independently from ${\mathcal N}(0, 1)$, $\{ \pm 1, \pm 3\}$ and $\{0, 255\}$, respectively, and their support indices are chosen at random.

For comparative purpose, the following representative approaches are included in our simulation: 
\begin{itemize}
\item BP~\cite{chen2001atomic};
\item OMP~\cite{Pati1993OMP};
\item PreOMP, i.e.,  the OMP algorithm preconditioned by Tsiligianni {\it et al.}~\cite{Tsiligianni2015preconditionofIUNTF};
\item mOLS~\cite{wang2017MOLS}, where the selection parameter  $s$ is chosen from $\{3, 5\}$ because $s \leq \min \{ K, \frac{m}{K}\}$;
\item PmOLS, where $s \in \{3, 5\}$. 
\end{itemize}

%

%

%

Fig.~\ref{fig1:Gauss}--\ref{fig1:POS} shows the performance comparison of different approaches for the recovery of sparse Gaussian signals,  sparse PAM signal and  sparse two-valued signals, respectively. Overall, it can be observed that for all three types of signals under test, the critical sparsity of the PmOLS algorithm is uniformly higher than those of OMP, PreOMP and mOLS. Even when compared with BP, PmOLS   still exhibits very competitive performance.

\subsection{Experimental Results} 

In this section, we perform imaging experiments to evaluate the performance of PmOLS. In our experiments, we consider  the typical GISC system and compare the imaging quality of PmOLS with other methods. The optical setup of GISC is specified in Fig.~\ref{fig:setup}.  A light-emitting diode (LED) with wavelength $\lambda = 532 nm$ is used as the light source. Through the K\"ohler illumination lens, the light beam generated from LED is projected on the digital micromirror device (DMD). The DMD consists of $1024\times786$ pixels, each of which has size $13 um \times 13 um$. Only $252\times 252$ pixels located at the center of the DMD are used to generate the designed pattern, because the light intensity is more evenly distributed in this area.  After being modulated by the designed pattern of the DMD, the light beam is projected on the object through an emission lens, whose magnification is $2.7$. The light signal reflected from the object is collected and recorded on a bucket detector by a conventional imaging lens.

In our experiments, we consider three types of objects: i) digit ``3'', ii) digit ``7'' and iii) ``Tai-chi'' of size $9 mm \times 9 mm $, which are digitized as images of $28 \times 28$ pixels and denoted as $\mathbf{X}$.  Also, we consider random Gaussian matrices $ \mathbf{ \Psi }$ of $n = 784$ columns, whose entries are drawn {\it i.i.d.} from Gaussian distribution $\mathcal{N}(0,{1}/{m})$. Each row of $ \mathbf{ \Psi }$ is reshaped into $28 \times 28$ two-dimension (2-D) image as a designed pattern. Since these pattern matrices may have negative entries, which cannot be imported into the DMD directly, we employ an operation called ``non-negative lifting''. To be specific,  by adding a positive constant $c_0$ to each entry of  these pattern matrices, we lift them to be  non-negative ones.
Then, the detected signal on the bucket detector is the component-wise product of  the pattern matrices and the digitized images to be retrieved. 
Put formally, the lifted sampling matrix becomes
$
 \mathbf{\Psi}_0 = \mathbf{\Psi} + \mathbf{C}_0,
$
where $\mathbf{C}_0 \in \mathbb{R}_{> 0}^{m \times n}$ has all entries being $c_0$'s.  
The detected signal on the bucket detector is given by
\begin{equation}
\mathbf{y}_0  = \mathbf{\Psi}_0 \mathbf{x}. 
\end{equation}


In GI, the object's information is retrieved from the fluctuations $\mathbf{ y }_0 - \langle \mathbf{ y }_0 \rangle$ and $\mathbf{\Psi}_0 - \langle \mathbf{\Psi}_0 \rangle$,
where the vector $ \langle \mathbf{ y }_0 \rangle $ has entries of the same value being the mean value of the entries in $\mathbf{y}$, while each column of $\langle \mathbf{\Psi}_0 \rangle$ has identical entries that equal to the mean value of the column of $ \mathbf{\Psi}_0 $. Specifically, the object $\mathbf{x}$ is recovered by correlating the intensity fluctuations of light fields, namely
\begin{equation}
   \hat{ \mathbf{x} } = (\mathbf{\Psi}_0 - \langle \mathbf{\Psi}_0 \rangle)^T (\mathbf{ y }_0 - \langle \mathbf{ y }_0 \rangle). \label{eq:28}
\end{equation}
Some modified versions of GI, such as  DGI~\cite{Ferri2010DGI} and  PGI~\cite{Zhang2014PGI}, suggest modifications on the correlation in~\eqref{eq:28} to improve the imaging quality. 
Our experiments include GI, DGI~\cite{Ferri2010DGI},  PGI~\cite{Zhang2014PGI}, mOLS~\cite{wang2017MOLS}, gradient projection for sparse reconstruction (GPSR)~\cite{Figueiredo2008GPSR} and BP~\cite{chen2001atomic}. 
%
%
%
Fig.~\ref{fig:experimental results} shows the imaging results of those approaches under different sampling rates, where the number of samples is labeled in the leftmost column, while the ground truth is placed in the right-most  column. It can be observed that for all objects under test, the PmOLS algorithm outperforms other methods in the imaging  quality. 
%

Moreover, to quantitatively evaluate the imaging quality of PmOLS, we employ the peak signal-to-noise ratio (PSNR) as the performance metric, which is defined as
\begin{equation}
 \rm{PSNR} := 20 \log_{10} \left(  \frac{255}{\sqrt{\rm{MSE} ( \hat{\mathbf{X}}, \mathbf{X} ) }}  \right),
\end{equation}
 where $\rm{MSE} ( \hat{\mathbf{X}}, \mathbf{X} )$ denotes the mean squared error between the reconstructed image $\hat{\mathbf{X}}$ and ground truth $\mathbf{X}$, 
 \begin{equation}
 \rm{MSE}   ( \hat{\mathbf{X}}, \mathbf{X} ) := \frac{1}{28 \times 28} \sum \limits_{i = 1}^{28} \sum \limits_{j = 1}^{28} ( \hat{ \mathbf{X} }_{i,j} - \mathbf{X}_{i,j} )^2.
 \end{equation}
A larger PSNR usually implies better reconstruction quality of images. The PNSR of recovered images (in dB) via different approaches are given in Table~\ref{tab:my_label}. Overall, it can be observed that the experimental results well match those in our numerical simulations. In particular,  the PmOLS algorithm performs comparably to BP and outperforms all the other approaches. Also, it can be observed that PmOLS improves the PSNR of reconstructed images by $1$dB or so over that of the classic mOLS algorithm, which clearly demonstrates the advantage of our PIP method.

\renewcommand\arraystretch{2}
\setlength{\arrayrulewidth}{.75pt}
\begin{table*}[h]
\caption{Comparison of the PSNR results of three objects (in $\textbf{dB}$) imaged by different algorithms.}
\centering
\begin{tabular}{|c|c|c|c|c|c|c|c|c|}
\hline
 Object & Sampling Size & GI & PGI~\cite{Zhang2014PGI} & DGI~\cite{Ferri2010DGI} & mOLS (${s=3}$)~\cite{wang2017MOLS} & GPSR~\cite{Figueiredo2008GPSR} & BP~\cite{chen2001atomic} & PmOLS (${s=3}$) \\
\hline  \hline 
digit ``$3$'' & $360$ & $20.48$ & $20.66$ & $20.67$ & $21.32$ & $21.82$ & $22.59$ & ${\bf 22.84}$ \\
\hline 
digit ``$7$'' & $420$ & $20.18$ & $20.33$ & $20.38$ & $21.77$ & $21.37$ & $22.40$ & ${\bf 22.83}$ \\
\hline 
``Tai-chi'' & $410$ & $20.14$ & $20.81$ & $20.17$ & $22.27$ & $21.52$ & $22.60$ & ${\bf 23.23}$ \\
\hline 
\end{tabular}
\label{tab:my_label}
\end{table*}

\section{DISCUSSION} \label{sec:VI}
In this section, we shall discuss some issues that arise from our analysis, simulation and experimental results.

\subsection{Related to Frame Theory}

The PIP method can be understood from the frame perspective. The theory of frame was first introduced by Duffin and Schaeffer~\cite{Duffin&Schaeffer1952frames} in the early 1950's and has evolved into a state-of-the-art signal processing tool in recent decades. Specific types of tight frames are a natural generalization of orthogonal bases in sparse and redundant representations. Interestingly, it can be shown that the preconditioned sampling matrix $\mathbf{P} \mathbf{\Psi}$ with $\mathbf{P} = \mathbf{\Psi}^{T} ( \mathbf{\Psi} \mathbf{\Psi}^{T} ) ^{-1}$ is a Parseval tight frame. Specifically, $\mathbf{P}\mathbf{\Psi}$ can be given by
 \begin{eqnarray}
 \mathbf{P} \mathbf{\Psi} 
& \hspace{- 2mm} \overset{\eqref{eq:itit}}{=} & \hspace{- 2mm} \mathbf{V} \hspace{.25mm} \text{diag} \big( [\underbrace{1,\cdots, 1}_{m},\underbrace{0,\cdots, 0}_{n-m} ] \big) \mathbf{V}^T  \nonumber\\
 & \hspace{- 2mm} = & \hspace{- 2mm} \left[ \mathbf{V}_m, \overline{\mathbf{V}}_m \right]
 \hspace{-1mm} 
 \left[ \begin{array}{*{20}{c}}
\hspace{-1mm} \mathbf{1}_{m \times m} & \mathbf{0}_{m \times (n - m)} \hspace{-1mm} \\
\hspace{-1mm} \mathbf{0}_{(n - m) \times m} & \mathbf{0}_{(n - m) \times (n - m)} \hspace{-1mm} 
\end{array} \right] 
 \hspace{-1mm}
 \left[ \begin{array}{*{20}{c}}
\hspace{-1mm} \mathbf{V}_m^T  \hspace{-1mm} \\
\hspace{-1mm} \overline{\mathbf{V}}_m^T \hspace{-1mm}
\end{array} \right] \nonumber \\
 & \hspace{- 2mm} = & \hspace{- 2mm} \mathbf{V}_m \mathbf{V}_m^T, \label{preconditioned sampling matrix0}
 \end{eqnarray}
where $\mathbf{V}_m   \in \mathbb{R}^{n \times m}$ and $\overline{ \mathbf{V} }_m   \in \mathbb{R} ^{n \times (n-m)}$ denote the restrictions of matrix $\mathbf{V}$ to its first $m$ columns and the remaining $n-m$ ones, respectively. 
Thus,
\begin{equation}
\label{gram}
( \mathbf{P} \mathbf{\Psi} )^T \mathbf{P} \mathbf{\Psi} =  (\mathbf{V}_m \mathbf{V}_m^T )^{T} \mathbf{V}_m \mathbf{V}_m^T   = \mathbf{V}_m \mathbf{V}_m^T.
\end{equation}
Since the nonzero eigenvalues of matrix $\mathbf{V}_m \mathbf{V}_m^{T}$ are all ones, by the properties of tight frame~\cite{Bhatt2018incoherentfiniteframe} the preconditioned sampling matrix $\mathbf{P} \mathbf{\Psi}$ is a Parseval tight frame.

\begin{figure*}[t]
\centering
\subfigure[Mutual coherence as a function of sampling rate]
{\hspace{-6mm} \includegraphics[width = 85mm, height = 85mm] {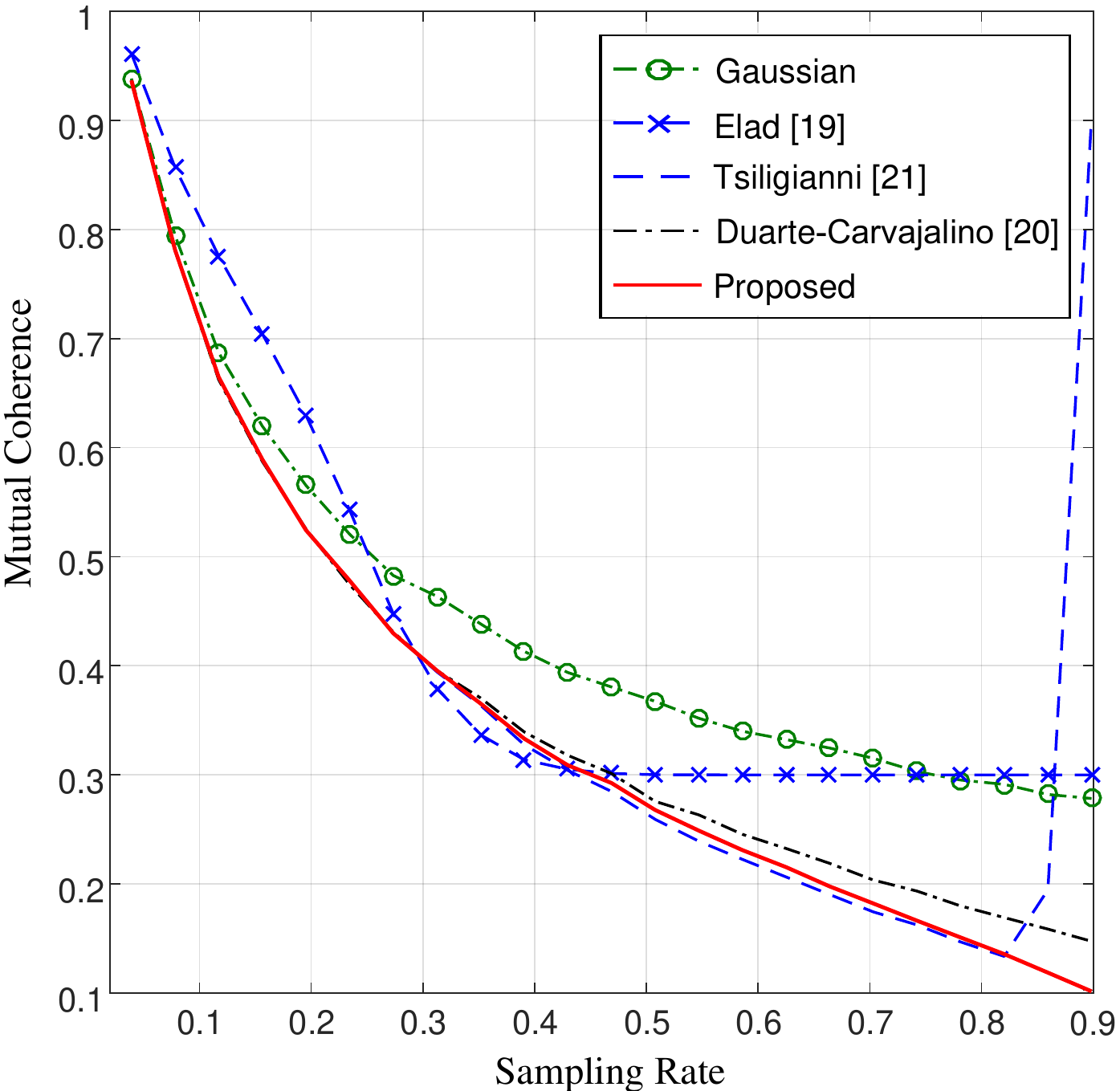}
\label{fig:mucomparsion} \hspace{5mm}}
\subfigure[MSE as a function of SNR]
{\includegraphics[width = 85mm, height = 85mm] {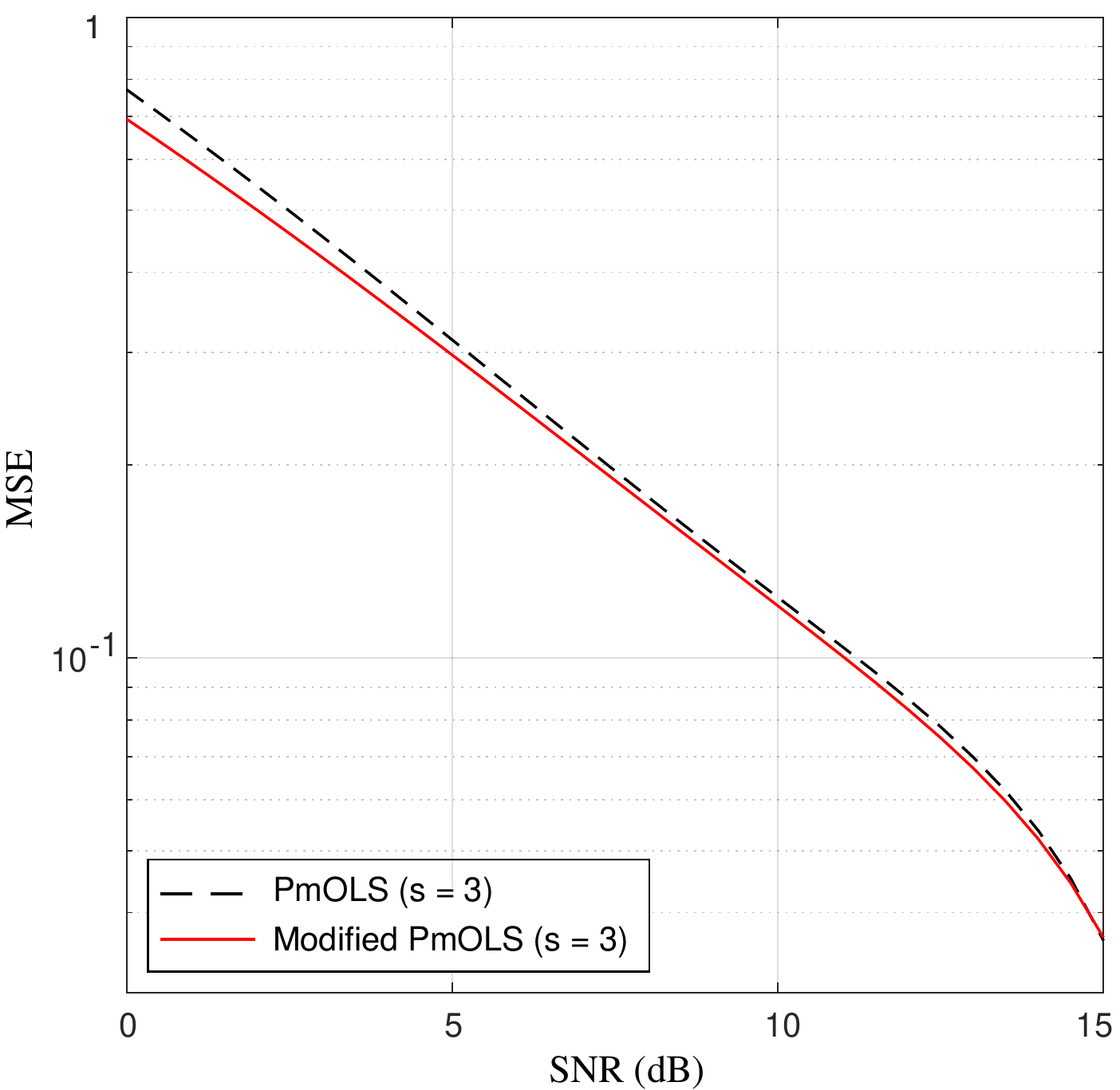}
\label{fig:SNR_modified_pmols}}
\caption{Simulation results: (a) Mutual coherence as a function of sampling rate; (b) MSE for recovering $50$-sparse Gaussian signals.}
\end{figure*}

\subsection{What If We Precondition $\mathbf{\Psi}$ Twice?}

So far, we have demonstrated from simulation results that the mutual coherence of sampling matrices can be improved via the proposed PIP method. A natural question is that whether the mutual coherence can be further reduced by preconditioning the matrices again.
The following theorem aims to give a negative answer to this question. 

\begin{theorem}
Let $\mathbf{\Psi} \in \mathbb{R}^{m \times n}$ be a sampling matrix with $m < n$ and let $\mathbf{P}$ be a preconditioner produced by PIP. Then, the mutual coherence of the preconditioned sampling matrix $\mathbf{P} \mathbf{\Psi}$ cannot be further reduced by applying the PIP method again.
\end{theorem}
\begin{IEEEproof}
As show in~\eqref{preconditioned sampling matrix0}, the preconditioned sampling matrix via PIP is
\begin{equation} \label{eq:44}
    \mathbf{P} \mathbf{\Psi} = \mathbf{V}_m \mathbf{V}_m^T.
\end{equation}
To see if the mutual coherence of matrix $\mathbf{P} \mathbf{\Psi}$ can be further reduced, we apply the PIP method again. Since $\mathbf{P}\mathbf{\Psi}$ is an $n \times n$ matrix of  rank $m$  ($< n$), by~Proposition~\ref{prop:P} the preconditioner for  $\mathbf{P}\mathbf{\Psi}$ can be given by
\begin{eqnarray}
 \mathbf{P}_{\mathbf{P}\mathbf{\Psi}} & \hspace{-2mm} = & \hspace{-2mm}  \mathbf{V}   \text{diag} \bigg(   \bigg[    \frac{1}{\sigma_1 (\mathbf{P}\mathbf{\Psi})},  \cdots , \frac{1}{\sigma_m (\mathbf{P}\mathbf{\Psi})} ,\underbrace{0,   \cdots  , 0}_{n - m} \bigg]   \bigg)   \mathbf{U}^T  \nonumber \\
& \hspace{-2mm} \overset{(a)}{=} & \mathbf{V}_m \mathbf{V}_m^T,
\end{eqnarray}
where (a) is because $\mathbf{P} \mathbf{\Psi} = \mathbf{V}_m \mathbf{V}_m^T$ so that its singular values $\sigma_i (\mathbf{P}\mathbf{\Psi})$, $i = 1, \cdots, m$, are all ones.  

Therefore,  
\begin{equation}
 \mathbf{P}_{\mathbf{P}\mathbf{\Psi}}  \mathbf{P} \mathbf{\Psi}  = \mathbf{V}_m \mathbf{V}_m^T \mathbf{V}_m \mathbf{V}_m^T = \mathbf{V}_m \mathbf{V}_m^T =  \mathbf{P} \mathbf{\Psi},
\end{equation}
which completes the proof.
\end{IEEEproof}

\subsection{Comparison of Preconditioning Methods}

We empirically compare the performance of  PIP with existing methods, including Elad~\cite{Elad2007optimizedprojection}, Tsiligianni {\it et al.}~\cite{Tsiligianni2015preconditionofIUNTF} and Duarte-Carvajalino {\it et al.}~\cite{duarte2009Sparsifyingdictionaryoptimization}. Fig.~\ref{fig:mucomparsion} shows the comparison of  mutual coherence of random Gaussian matrices after applying different preconditioning methods. It can  be observed that for all testing methods except Elad's~\cite{Elad2007optimizedprojection}, the mutual coherence of preconditioned matrices is uniformly better than the original Gaussian case.  To be specific, the proposed PIP method performs similarly with the methods of Tsiligianni {\it et al.}~\cite{Tsiligianni2015preconditionofIUNTF} and Duarte-Carvajalino {\it et al.}~\cite{duarte2009Sparsifyingdictionaryoptimization} in low sampling rates and performs the best in the high sampling rates. Elad's method~\cite{Elad2007optimizedprojection} is effective only when the sampling rate is below $0.4$. Moreover, the method of Tsiligianni {\it et al.}~\cite{Tsiligianni2015preconditionofIUNTF} works pretty well  in the low sampling rate region but degrades sharply when the sampling rate goes beyond $0.8$. This is perhaps because this method utilizes the alternative projection to solve a non-convex optimization problem, which may not converge in some cases. 
%
%
Overall, we observe that the proposed PIP method is competitive compared to the existing methods.

%
%

\subsection{Alternative to Non-negative Lifting}
In our experimental section, we have employed a non-negative lifting to cope with the non-negative constraint  of sampling matrices (i.e., $\mathbf{\Psi}_{ij} \geq 0, \forall i,j$). In the worst case, however, this operation may severely affect the mutual incoherence of sampling matrices, leading to potential risk to the imaging quality. To deal with this issue, we provide an alternative operation called non-negative subtraction.  The key idea is to use a non-negative matrix to sample the input signal $\mathbf{x}$, but use a different matrix (i.e., the original sampling matrix $\mathbf{\Psi}$) for the sequent reconstruction, which can be realized in the following steps.

\begin{enumerate}[i)]

\item Since the  sampling matrix $\bf{\Psi}$ (e.g., the random Gaussian matrix considered in this paper) does not  necessarily have all non-negative entries, we decompose it into a subtraction of two non-negative matrices:
\begin{equation}
 \bf{ \Psi} = \bf{ \Psi}_+ -  \bf{ \Psi}_-. \label{eq:phi}
\end{equation}
For instance, 
\begin{equation} 
\hspace{-.5mm}    \overbrace{\left[ \begin{array}{ccc}
     1           & 1                     & 1 \vspace{-2mm}  \\ 
     -1 &  -1      & -1
    \end{array} \right]}^{{\bf{ \Psi}}} 
     \hspace{-.25mm} = \hspace{-.25mm}
        \overbrace{\left[ \begin{array}{ccc}
     1           & 1                   & 1  \vspace{-2mm}  \\
     0 &  0      & 0 
    \end{array} \right]}^{{\bf{ \Psi}}_+} 
    \hspace{-.25mm} -  \hspace{-.25mm}
        \overbrace{\left[ \begin{array}{ccc}
     0           & 0                      & 0 \vspace{-2mm}   \\
     1 &  1      & 1  
    \end{array} \right]}^{{\bf{ \Psi}}_-} \hspace{-.25mm}. \nonumber
\end{equation}

\item The matrices $ {\bf{ \Psi}}_+$ and $ {\bf{ \Psi}}_-$  are used to physically acquire the samples of  $\mathbf{x}$ as
\begin{equation}
    \left[ \begin{array}{ccccc}
    & &  {\bf{ \Psi}}_+ & &  \\
    \hdashline
    & &  {\bf{ \Psi}}_-  & &
    \end{array} \right] 
    \left[ \begin{array}{c}
      \\
    \hspace{-1mm}  \mathbf{x}  \hspace{-1mm}  \\
    \\
    \end{array} \right] 
    = \left[ \begin{array}{c}
    \hspace{-1mm} {\bf{y}}_+   \hspace{-2mm} \\
     \hdashline 
     \hspace{-1mm} {\bf{y}}_-  \hspace{-2mm}
    \end{array} \right],
\end{equation}

\item  Using these samples,  we obtain the desired samples:
\begin{equation}
{\bf{y}} = {\bf{y}}_+ - {\bf{y}}_- =  (  {\bf{ \Psi}}_+ -   {\bf{ \Psi}}_- )  {\bf{x}} \overset{\eqref{eq:phi}}{=} \mathbf{ \Psi} \mathbf{x},
\end{equation}
from which we can recover $\mathbf{x}$. 
\end{enumerate}

The non-negative subtraction has theoretical advantages over the non-negative lifting as it does not affect the mutual coherence of $ {\bf{ \Psi}}$, which is no doubt beneficial to the image reconstruction. However, it should be pointed out that this theoretical benefit is essentially at the cost of acquiring double samples (i.e., $\bf{y}_+$ and ${\bf{y}}_-$). Considering i) that the cost of acquiring more samples could be very expensive in practice and ii) that these two operations generally lead to very similar performance as confirmed by our numerical simulations, we still suggest to use the non-negative lifting in the experiments.  

%
%
%

\subsection {The Noisy Case}
In practical GISC systems, the detection noise is unavoidable and may even be enlarged by preconditioning, which in turn becomes detrimental to the signal reconstruction. To address this issue, we propose a regularization method to minimize the weighted sum of the
objectives:
\begin{equation} \label{min:P1}
 \mathop {\min} \limits_{\mathbf{P}} {\left\| \mathbf{P} \mathbf{\Psi} - \mathbf{I} \right\|_F} + \lambda \|\mathbf{P}\|_F,
\end{equation}
for  some $\lambda  > 0$ related to the level of detection noise. One can show that this problem has a closed-form solution:
 \begin{equation}
      \mathbf{P}_{\lambda} =  \mathbf{\Psi}^T \left( \mathbf{\Psi} \mathbf{\Psi}^T + \lambda \mathbf{I} \right) ^{-1}, \label{eq:eq32}
\end{equation}  
which we call the modified PIP method.

To show the effectiveness of this solution, we empirically test the MSE performance for the recovery of sparse signals. In our simulation,  we follow the simulation settings in~\cite{wang2017MOLS}, in which the system model is given by
\begin{equation}
 \mathbf{y}_0 = \mathbf{\Psi} \mathbf{x} + \mathbf{v}
 \end{equation} 
and  the noise vector $ \mathbf{v} $ has elements generated {\it i.i.d.} from $ \mathcal{ N } ( 0 , \frac{K} {m} 10 ^ { - \frac{ \text{ SNR } } { 10 } } ) $, where 
\begin{equation}
 \text{SNR} := 10 \log_{10} \frac{ \| \mathbf{\Psi} \mathbf{ x } \|_2^2} { \| \mathbf{v} \|_2^2}
\end{equation}
is the signal to noise rate (SNR).  We consider the sampling matrix $\mathbf{\Psi} \in \mathbb{R}^{ 128 \times 256 }$ to have entries drawn {\it i.i.d.} from $\mathcal{N}(0,\frac{1}{128})$ and the input signal $\mathbf{x}$ to be $50$-sparse and have entries drawn {\it i.i.d.} from $\mathcal{N}(0,1)$. 
In Fig.~\ref{fig:SNR_modified_pmols},  the MSE performance of the modified PmOLS is shown as a function of SNR. The benchmark performance of PmOLS is plotted as well. 
Clearly it can be observed that the MSE of the modified PmOLS algorithm  (i.e., the modified PIP method followed by mOLS)  is uniformly better than that of  PmOLS, especially at low SNR's. Therefore, the modified PIP method can be used to improve the reconstruction quality of sparse signals in the noisy scenario.

%

\section{CONCLUSION}\label{sec:VII}

In this paper, with an aim of obtaining faithful imaging quality via GISC, we have 
proposed an algorithm called PmOLS, which consists of two parts: i) the PIP method and ii) the mOLS algorithm. Through theoretical analysis, we have shown that the PIP method can well bound the mutual coherence of sampling matrices with high probability. The analysis of the PIP method is potentially useful for the application to other sparsity-constrained optimization problems, such as sparse phase retrieval~\cite{Gerchberg1972GS,candes2015PRWF}. 
In the second part of  PmOLS  is the classic mOLS algorithm. For this algorithm,  we have built a mutual coherence based condition, which, in conjunction with the probabilistic guarantee of the PIP method, leads to an overall recovery condition for the PmOLS algorithm. Therefore,  the choice of $m \geq c K^2 \log (n/\epsilon)$ is sufficient to ensure the success probability exceeding $1 - \epsilon$.

We would like to mention a technical limitations in our analysis. Specifically, by introducing a regularization term, we have proposed a closed-form solution of the preconditioner under noise, which, however, only brings in marginal improvement over the PIP method, as shown in Fig.~\ref{fig:SNR_modified_pmols}. Designing a more effective preconditioner that is robust to noise and deriving analytical results demonstrating the effectiveness can be of great significance. 
Our future work will be directed towards this avenue.

%
%

\appendices

\section{Proof of Lemma \ref{sigma of Gauss matrix} } \label{app:0}
\begin{IEEEproof}

Let $\mathbf{\Psi} \in {\mathbb{R}^{m \times n}}$ be a random Gaussian matrices with entries ${\varphi _{i,j}}\mathop  \sim \limits^{i.i.d} \mathcal{N} \left( 0, {1}/{m} \right)$ and let 
\begin{equation}
\varphi_{i,j}' = \sqrt{ {m}/{n}} \hspace{.75mm}\varphi_{i,j},~\forall i,j.
\end{equation}
Then,  $\mathbf{\Psi}' \in {\mathbb{R}^{n \times m}}$ 
has entries ${\varphi _{i,j}'}\mathop  \sim \limits^{i.i.d} \mathcal{N} \left( 0, {1}/{n} \right)$.
Also, let $\sigma_1(\mathbf{\Psi}) \ge \cdots \ge \sigma_m (\mathbf{\Psi})$ and $\sigma_1(\mathbf{\Psi}') \ge \cdots \ge \sigma_m (\mathbf{\Psi}')$ be the singular values of matrices $\mathbf{\Psi}$ and $\mathbf{\Psi}'$, respectively.  
Then, 
\begin{equation} \label{A.2}
    \sigma_i(\mathbf{\Psi}) = \sqrt{ {n}/{m}} \hspace{.75mm}\sigma_i(\mathbf{\Psi}'),~~ \forall i \in \{ 1, \cdots, m \}
\end{equation}

In~\cite[Section 4.1]{candes2006sigmagaussianmatrix}, it has been shown that for any constant $\varepsilon \geq 0$, $\sigma_1(\mathbf{\Psi}')$ and $\sigma_m(\mathbf{\Psi}')$ satisfy
\begin{subequations}
\begin{align}
     &  \Pr \big\{ \sigma_1 (\mathbf{\Psi}')  \geq  1 + \sqrt{ {m}/{n}} + \varepsilon  \big \}  \leq e^ {- n \varepsilon^2 / 2}, \label{ A3.a }\\
     &  \Pr \big \{ \sigma_m  (\mathbf{\Psi}') \hspace{-.25mm}  \leq  \hspace{-.25mm}  1 \hspace{-.25mm} - \hspace{-.25mm} \sqrt{ {m}/{n}} - \varepsilon  \big \}  \leq e^ {- n \varepsilon^2 / 2}.\label{ A3.b }  
\end{align} 
\end{subequations} 
Using~\eqref{A.2},~\eqref{ A3.a } and~\eqref{ A3.b }, we can build the lower and upper bounds for $\sigma_1(\mathbf{\Psi})$ and $\sigma_m(\mathbf{\Psi})$, respectively. Specifically, we have
\begin{eqnarray}
 \Pr \{ \sigma_1 (\mathbf{\Psi})  \geq  t \}  & \hspace{-2mm} \overset{\eqref{A.2}}{=} & \hspace{-2mm} \Pr  \big \{ \sigma_1 (\mathbf{\Psi}')  \geq  \sqrt{ {m}/{n} } \hspace{0.75mm} t  \big \} \nonumber \\
  & \hspace{-2mm} \overset{\eqref{ A3.a }}{ \leq }& \hspace{-2mm} e^ {- n ( \sqrt{ {m}/{n} } \hspace{.5mm} t - \sqrt{ {m}/{n}} -1 )^2 / 2} \nonumber \\
  & \hspace{-2mm} \leq & \hspace{-2mm} e^ {- m ( t - \sqrt{ {n}/{m} } -1 )^2 / 2}.
\end{eqnarray}
Let $\varepsilon = t - \sqrt{ {n}/{m}} -1$. Then, 
\begin{equation}
    \Pr \big\{ \sigma_1 (\mathbf{\Psi})  \geq  \sqrt{ {n}/{m} } + 1 +  \varepsilon  \big \} \leq e^ {- m \varepsilon^2 / 2}.
\end{equation}
Likewise, $\sigma_m (\mathbf{\Psi})$ can be upper bounded as
\begin{eqnarray}
 \Pr \{ \sigma_m (\mathbf{\Psi})  \leq  t \} &  \hspace{-2mm}  \overset{\eqref{A.2}}{=} &  \hspace{-2mm}  \Pr \big \{ \sigma_m' (\mathbf{\Psi}')  \leq  \sqrt{ {m}/{n} } \hspace{.75mm} t \big \} \nonumber \\
  &  \hspace{-2mm}  \overset{\eqref{ A3.b }}{\leq}&  \hspace{-2mm}  e^ {- n ( 1 -  \sqrt{ {m}/{n} } - \sqrt{ {m}/{n} } t )^2 / 2} \nonumber \\
  & \hspace{-2mm}  \leq &  \hspace{-2mm}  e^ {- m (\sqrt{ {n}/{m} } - 1 - t )^2 / 2}.
\end{eqnarray}
By letting $\varepsilon = \sqrt{ {n}/{m}} - 1 - t $, we have
\begin{equation}
    \Pr \big \{ \sigma_m (\mathbf{\Psi})  \leq  \sqrt{ {n}/{m}} - 1 -  \varepsilon \big \} \leq  e^ {- m \varepsilon^2 / 2}.
\end{equation}
The proof is thus complete.
\end{IEEEproof}

\section{Proof of Proposition \ref{prop:P}} \label{app:P}

\begin{IEEEproof}

we consider the three cases for solving~\eqref{min:P}. 
 \begin{enumerate}[i)]
 \item If  $r = m$, then matrix $\mathbf{\Psi}$ has  full row rank. One can show that problem \eqref{min:P} has the closed-form solution:
     \begin{equation} \label{eq:gagaga1}
      \mathbf{P} = \mathbf{\Psi}^T \left( \mathbf{\Psi} \mathbf{\Psi}^T \right) ^{-1}.
      \end{equation} 
 Specifically, let $\mathbf{U} \mathbf{\Sigma} \mathbf{V}^T$ be the singular value decomposition (SVD) of $\mathbf{\Psi}$ where $\mathbf{\Sigma} = [\mathbf{\Sigma}_m ~\mathbf{0}]$ and $\mathbf{\Sigma} _m$ is a diagonal matrix with diagonal elements being the nonzero singular values of $\mathbf{\Psi}$ in descending order. Then, \eqref{eq:gagaga1} implies 
      \begin{eqnarray} \label{eq:vvv1}
            \mathbf{P} \mathbf{\Psi} &=&  \mathbf{V} \mathbf{\Sigma}^T  \mathbf{U}^T 
    \left (  \mathbf{U} \mathbf{\Sigma} \mathbf{V}^T  \mathbf{V} \mathbf{\Sigma}^T  \mathbf{U}^T  \right)^{-1} \mathbf{U} \mathbf{\Sigma} \mathbf{V}^T  \nonumber \\
    &=& \mathbf{V} \hspace{.25mm} \text{diag} \big( [\underbrace{1,\cdots, 1}_{m},\underbrace{0,\cdots, 0}_{n-m} ] \big) \mathbf{V}^T, 
      \end{eqnarray}
where $\text{diag}(\mathbf{u})$ returns a square diagonal matrix with the elements of vector $\mathbf{u}$ on the main diagonal. Clearly matrix $\mathbf{P} \mathbf{\Psi}$ in~\eqref{eq:vvv1} has the minimum Frobenius norm distance to the identity matrix $\mathbf{I} \in \mathbb{R}^{n \times n}$.

 \item If $ r = n $, then matrix $\mathbf{\Psi}$ has full column rank. In a similar way, one can also show that problem \eqref{min:P} has the closed-form solution:
     \begin{equation} \label{eq:gagaga2}
      \mathbf{P} = \left( \mathbf{\Psi}^T \mathbf{\Psi} \right) ^{-1} \mathbf{\Psi}^T. 
      \end{equation} 
      To be specific, let $\mathbf{U} \mathbf{\Sigma} \mathbf{V}^T$ be the SVD of $\mathbf{\Psi}$ where $\mathbf{\Sigma} = [\mathbf{\Sigma}_n ~\mathbf{0}]^T$ and $\mathbf{\Sigma} _n$ is a diagonal matrix with diagonal elements being the nonzero singular values of $\mathbf{\Psi}$ in descending order. 
      Then it follows from \eqref{eq:gagaga2} that
      \begin{eqnarray} \label{eq:vvv2}
            \mathbf{P} \mathbf{\Psi} &=&  
    \left ( \mathbf{V} \mathbf{\Sigma}^T  \mathbf{U}^T  \mathbf{U} \mathbf{\Sigma} \mathbf{V}^T   \right)^{-1} \mathbf{V} \mathbf{\Sigma}^T  \mathbf{U}^T \mathbf{U} \mathbf{\Sigma} \mathbf{V}^T   \nonumber \\
    &=& \mathbf{V} \hspace{.25mm} \text{diag} \big( [\underbrace{1,\cdots, 1}_{n}] \big) \mathbf{V}^T,
      \end{eqnarray}
which  is equal to the identity matrix $\mathbf{I} \in \mathbb{R}^{n \times n}$. 

  \item If $r < \min\{ m, n\}$, then matrix $\mathbf{\Psi}$ is neither full column nor full row rank.  Nevertheless, one can still show that problem \eqref{min:P} has the closed-form solution:
     \begin{equation} \label{eq:gagaga3} 
      \mathbf{P} = \mathbf{V} \left[ \begin{array}{*{20}{c}}
\mathbf{\Sigma}_r ^{-1}& \mathbf{0}_{r \times (n - r)} \\
\mathbf{0}_{(m - r) \times r} & \mathbf{0}_{(m - r) \times (n - r)} \hspace{-1mm} 
\end{array} \right]  \mathbf{U}^T,
      \end{equation}  
    where $\mathbf{\Sigma} _r$ is a diagonal matrix with diagonal elements being $r$  nonzero singular values of $\mathbf{\Psi}$ in descending order. 
   Indeed, let $\mathbf{U} \mathbf{\Sigma} \mathbf{V}^T$ be the SVD of $\mathbf{\Psi}$ where
   \begin{equation}
   \mathbf{\Sigma} = \left[ \begin{array}{*{20}{c}}
\mathbf{\Sigma}_r & \mathbf{0}_{r \times (n - r)} \\
\mathbf{0}_{(m - r) \times r} & \mathbf{0}_{(m - r) \times (n - r)} \hspace{-1mm} 
\end{array} \right] .
   \end{equation} 
   Then, \eqref{eq:gagaga3} implies 
      \begin{eqnarray} \label{eq:vvv3}
            \mathbf{P} \mathbf{\Psi} &=&  \mathbf{V} \left[ \begin{array}{*{20}{c}}
\mathbf{\Sigma}_r ^{-1}& \mathbf{0}_{r \times (n - r)} \\
\mathbf{0}_{(m - r) \times r} & \mathbf{0}_{(m - r) \times (n - r)} \hspace{-1mm} 
\end{array} \right]  \mathbf{U}^T  \nonumber\\
&& \mathbf{U}\left[ \begin{array}{*{20}{c}}
\mathbf{\Sigma}_r  & \mathbf{0}_{r \times (n - r)} \\
\mathbf{0}_{(m - r) \times r} & \mathbf{0}_{(m - r) \times (n - r)} \hspace{-1mm} 
\end{array} \right] \mathbf{V}^T    \nonumber \\
    &=& \mathbf{V} \hspace{.25mm} \text{diag} \big( [\underbrace{1,\cdots, 1}_{r}, \underbrace{0,\cdots, 0}_{n - r}] \big) \mathbf{V}^T,
      \end{eqnarray}
 which has the minimum Frobenius-norm distance to the identity matrix $\mathbf{I} \in \mathbb{R}^{n \times n}$.
 \end{enumerate}
 
In summary, the preconditioned sampling matrix $\mathbf{P}\mathbf{\Psi}$ satisfies
\begin{equation}
    \mathbf{P} \mathbf{\Psi} = \mathbf{V} \hspace{.25mm} \text{diag} \big( [\underbrace{1,\cdots, 1}_{\text{rank}(\mathbf{\Psi})},\underbrace{0,\cdots, 0}_{n - \text{rank}(\mathbf{\Psi})} ] \big) \mathbf{V}^T,
\end{equation}
which completes the proof.
\end{IEEEproof}

\section{Proof of Theorem \ref{theorem: mu of preconditioned matrix}}\label{app:1}
\begin{IEEEproof}

%
%

In a nutshell, the proof of Theorem \ref{theorem: mu of preconditioned matrix} consists of two parts. Firstly, we build the relationship between  $\mu (\mathbf{P \Psi})$ and $\mu (\mathbf{\Psi})$, which allows to characterize the lower bound of $\Pr (\mu (\mathbf{P \Psi}) \leq \eta)$ in terms of both the matrix $\mu (\mathbf{\Psi})$ and the singular values of $\mathbf{\Psi}$.  Some of our proof follows along a  similar line as in the recent work~\cite{Chen2018preomp}.  Secondly, we further estimate the lower bound of $\Pr (\mu (\mathbf{P \Psi}) \leq \eta)$ by incorporating the random Gaussian nature of $\mathbf{\Psi}$. 

\subsection{Relationship between $\mu (\mathbf{P \Psi})$ and $\mu (\mathbf{\Psi})$}
We first simplify the preconditioned sampling matrix $\mathbf{P} \mathbf{\Psi}$.  
Let $\mathbf{U} \mathbf{\Sigma} \mathbf{V}^T$ be the SVD of $\mathbf{\Psi}$,  where $\mathbf{\Sigma} = [\mathbf{\Sigma}_m ~\mathbf{0}_{m \times (n - m)}]$ and $\mathbf{\Sigma} _m = \text{diag} \big( [\sigma_1,\cdots, \sigma_m] \big)$ is a diagonal matrix with diagonal elements being the singular values of $\mathbf{\Psi}$ in descending order of their magnitudes. 
 Then,  
 \begin{eqnarray}
 \mathbf{P} \mathbf{\Psi} &  \hspace{- 2mm} \overset{ \eqref{eq:eq4} }{=} & \hspace{- 2mm} \mathbf{\Psi}^T \left( \mathbf{\Psi} \mathbf{\Psi}^T \right) ^{-1} \mathbf{\Psi} \nonumber\\
& \hspace{- 2mm} \overset{\eqref{eq:vvv1}}{=} & \hspace{- 2mm} \mathbf{V} \hspace{.25mm} \text{diag} \big( [\underbrace{1,\cdots, 1}_{m},\underbrace{0,\cdots, 0}_{n-m} ] \big) \mathbf{V}^T  \nonumber\\
 & \hspace{- 2mm} = & \hspace{- 2mm} \left[ \mathbf{V}_m, \overline{\mathbf{V}}_m \right]
 \hspace{-1mm} 
 \left[ \begin{array}{*{20}{c}}
\hspace{-1mm} \mathbf{1}_{m \times m} & \mathbf{0}_{m \times (n - m)} \hspace{-1mm} \\
\hspace{-1mm} \mathbf{0}_{(n - m) \times m} & \mathbf{0}_{(n - m) \times (n - m)} \hspace{-1mm} 
\end{array} \right] 
 \hspace{-1mm}
 \left[ \begin{array}{*{20}{c}}
\hspace{-1mm} \mathbf{V}_m^T  \hspace{-1mm} \\
\hspace{-1mm} \overline{\mathbf{V}}_m^T \hspace{-1mm}
\end{array} \right] \nonumber \\
 & \hspace{- 2mm} = & \hspace{- 2mm} \mathbf{V}_m \mathbf{V}_m^T, \label{preconditioned sampling matrix}
 \end{eqnarray}
where $\mathbf{V}_m  = [ \mathbf{v}_1, \cdots, \mathbf{v}_n ]^{T} \in \mathbb{R}^{n \times m}$ and $\overline{ \mathbf{V} }_m  = [ \overline{ \mathbf{v} }_1, \cdots, \overline{ \mathbf{v} }_n]^{T} \in \mathbb{R} ^{n \times (n-m)}$
 are the restrictions of matrix $\mathbf{V}$ to its first $m$ columns and the remaining $n-m$ ones, respectively.

From \eqref{preconditioned sampling matrix}, one can derive an upper bound of $\mu ( \mathbf{P} \mathbf{\Psi} )$ in terms of both $\mu ( \mathbf{\Psi} )$ and the singular values of $\mathbf{\Psi}$. To be specific, by the definition of mutual coherence, 
\begin{eqnarray}
    \mu ( \mathbf{P} \mathbf{\Psi} ) & \hspace{- 4mm} =& \hspace{- 3mm} \underset{1 \leq i <  j \leq n} {\max} \frac{ \big |  ( \mathbf{P} \mathbf{\Psi} )_i^{T}  ( \mathbf{P} \mathbf{\Psi} )_j  \big | } { \| ( \mathbf{P} \mathbf{\Psi} )_i  \|_2  \| ( \mathbf{P} \mathbf{\Psi} ) _j \|_2 } \nonumber \\
    & \hspace{- 4mm} \overset{(a)}{=} & \hspace{-3mm} \underset{1 \leq i <  j \leq n} {\max} \frac{ | ( \mathbf{V}_m   \mathbf{v}_i  )^{T}  \mathbf{V}_m \mathbf{v}_j | }{ \| \mathbf{V}_m \mathbf{v}_j \|_2 \| \mathbf{V}_m \mathbf{v}_j \|_2} \nonumber \\
    & \hspace{-3mm} = & \hspace{- 2mm} \mathop {\max }\limits_{1 \le i  <  j \le n} \frac{  |  \mathbf{v}_i   ^T  \mathbf{v} _j  | }   {  \| \mathbf{v} _i  \| _2   \|   \mathbf{v} _j  \|_2 } \nonumber \\
    &\hspace{-4mm} \overset{(b)}{=}& \hspace{-3mm} \frac{ | \mathbf{\psi}_i ^{T}  \mathbf{\Sigma}_m^{-2} \mathbf{\psi}_j  | }  { \| \mathbf{\Sigma}_m^{-1} \mathbf{\psi}_i \|_2 \hspace{.5mm} \| \mathbf{\Sigma}_m^{-1} \mathbf{\psi}_j \|_2 }  \nonumber \\
 &\hspace{-4mm} \overset{\text{Lemma}~\ref{wielandt inequality}}{\leq}& \hspace{-3mm} \frac{ ( \sigma_1^2 - \sigma_m^2 )  ( \| \mathbf{\psi}_i \|_2 \| \mathbf{ \psi }_j \|_2 )  + ( \sigma_1^2 + \sigma_m^2  ) | \mathbf{ \psi }_i^{T} \mathbf{\psi}_j |  } 
{ ( \sigma_1^2 + \sigma_m^2 )   ( \| \mathbf{ \psi}_i \|_2 \| \mathbf{\psi}_j \|_2 )  + ( \sigma_1^2 - \sigma_m^2  ) | \mathbf{\psi}_i^{T} \mathbf{\psi}_j |  } \nonumber \\
 &\hspace{-4mm} =& \hspace{-3mm} \frac{ \nu_m   + \frac{  |  \mathbf{\psi}_i   ^T  \mathbf{\psi} _j  | }   {  \| \mathbf{\psi} _i  \| _2   \|   \mathbf{\psi} _j  \|_2 } }
 {1 + \nu_m  \cdot \frac{  |  \mathbf{\psi}_i   ^T  \mathbf{\psi} _j  | }   {  \| \mathbf{\psi} _i  \| _2   \|   \mathbf{\psi} _j  \|_2 } } \nonumber\\
 &\hspace{-4mm} \overset{(c)}{\leq}& \hspace{-3mm} \frac{{\nu_m   + \mu \left( \mathbf{\Psi} \right)}}{{1 + \nu_m  \mu \left( \mathbf{\Psi} \right)}}   \leq   \nu_m + \mu \left( \mathbf{\Psi} \right)  \label{eq:A4},
\end{eqnarray}
 where $\nu_m := \frac{\sigma_1^2  - \sigma_m^2}{\sigma_1^2 + \sigma_m^2} \in (0, 1)$,
 $(a)$ is because $ ( \mathbf{P} \mathbf{\Psi} )_i = ( \mathbf{V}_m \mathbf{V}_m^{T})  _i  =  \mathbf{V}_m  \mathbf{v}_i$
and $\mathbf{V}_m^{T}\mathbf{V}_m = \mathbf{I}_{m \times m}$, (b) is because $\mathbf{\Psi} = \mathbf{U} \mathbf{\Sigma} \mathbf{V}^T = \mathbf{U} \mathbf{\Sigma}_m \mathbf{V}_m^T$  so that $ \mathbf{\psi}_i = \mathbf{U} \mathbf{\Sigma}_m \mathbf{v}_i$, and (c) is due to i) that the function $f(x) := \frac{\nu_m  + x}{ 1 + \nu_m x}$ is monotonically increasing in $x$ and ii) that 
 $$ \frac{  |  \mathbf{\psi}_i   ^T  \mathbf{\psi} _j  | }   {  \| \mathbf{\psi} _i  \| _2   \|   \mathbf{\psi} _j  \|_2 } \leq \max_{1 \leq i < j \leq n} \frac{  |  \mathbf{\psi}_i   ^T  \mathbf{\psi} _j  | }   {  \| \mathbf{\psi} _i  \| _2   \|   \mathbf{\psi} _j  \|_2 } \overset{\eqref{mutual coherence}}{=} \mu ( \mathbf{\Psi}). $$

Clearly \eqref{eq:A4} implies that
\begin{eqnarray} 
\Pr( \mu ( \mathbf{P} \mathbf{\Psi} ) \leq  \eta )  \geq   \Pr( \mu \left( \mathbf{\Psi} \right)  \leq \eta - \nu_m).  \label{eq:A88}
\end{eqnarray}

\subsection{Lower Bound of $\Pr( \mu ( \mathbf{P} \mathbf{\Psi} ) \leq  \eta )$}
One can see from \eqref{eq:A88} that to estimate a lower bound of $\Pr( \mu ( \mathbf{P} \mathbf{\Psi} ) \leq  \eta )$, it suffices to build that for $\Pr( \mu \left( \mathbf{\Psi} \right)  \leq \eta - \nu_m)$, where the parameter $\nu_m$ depends on the singular values of $\mathbf{\Psi}$ only. In  the following, 
we shall lower bound $\Pr( \mu \left( \mathbf{\Psi} \right)  \leq \eta - \nu_m)$ in three steps.  Before proceeding, define two events:
\begin{subequations}
\begin{align}
& \hspace{-2mm} E_1   \overset{\text{def}}{=}  \big\{ \sqrt{{n}/{m}} \hspace{-.5mm} - \hspace{-.5mm} 1 \hspace{-.5mm} - \hspace{-.5mm} \varepsilon   \hspace{-.5mm}
\leq    \hspace{-.5mm} \sigma_i  ( \mathbf{\Psi} )    \hspace{-.5mm} \leq   \hspace{-.5mm} \sqrt{{n}/{m}} \hspace{-.5mm} + \hspace{-.5mm} 1 \hspace{-.5mm} + \hspace{-.5mm} \varepsilon \big\} \\ \hspace{-.5mm} 
& \hspace{-2mm}  E_2  \overset{\text{def}}{=}  \bigg \{ \nu_m \leq \frac{ 2 \sqrt{{n}/{m}} \hspace{.5mm} (  1 + \varepsilon )} { {n}/{m} + ( 1 + \varepsilon )^2 }  \bigg \} \label{eq:iqivhg}
\end{align} 
\end{subequations}
for any constant $\varepsilon \geq 0$, where  $\sigma_i ( \mathbf{\Psi} ) $'s, $i = 1, \cdots, m$, are the singular values of the sampling matrix $\mathbf{\Psi}$ in descending order of their magnitudes.

\begin{enumerate}[i)]
\item {\bf Lower bound of $ \Pr  ( E_1 ) $:}

Since $\mathbf{\Psi}$ is a random Gaussian matrix with entries ${\varphi _{i,j}}\mathop  \sim \limits^{i.i.d.} \left( {0,{1}/{m}} \right)$, by Lemma~\ref{mutual coherence of Gauss matrix}  one has 
\begin{eqnarray}\label{probability:gaussian matrix svd}
 \Pr  ( E_1 )  
& \hspace{-2mm} \overset{(a)}{\geq} & \hspace{-2mm}
\Pr \big \{  \sqrt{{n}/{m}}  - 1 - \varepsilon   \leq  \sigma _i  ( \mathbf{\Psi} )  \big\}  \nonumber \\
& \hspace{-2mm} & \hspace{-2mm}  +  \Pr \big \{ \sigma _i  ( \mathbf{\Psi} )  \leq  \sqrt{{n}/{m}}  + 1  +   \varepsilon \big \}  -  1 \nonumber \\ 
& \hspace{-2mm}  \geq  & \hspace{-2mm}
\Pr \big \{ \sqrt{{n}/{m}}  - 1  - \varepsilon  \hspace{-0.5mm} \leq  \sigma _m ( \mathbf{\Psi} )   \big \}  \nonumber \\
& \hspace{-2mm}  & \hspace{-2mm}  +  \Pr \big \{  \sigma _1 ( \mathbf{\Psi} )   \leq \sqrt{{n}/{m}}  + 1  +  \varepsilon \big \}  -   1 \nonumber \\
& \hspace{-2mm} \overset{(b)}{ = } & \hspace{-2mm}  1 - 2 e^{ - { m \varepsilon ^2 }/{2} },   \label{eq:72ygygy}
\end{eqnarray}
where (a) uses the fact that $1 \geq \Pr(A \cup B) = \Pr(A) + \Pr(B) - \Pr(A \cap B)$ with events $A := \{ \sqrt{{n}/{m}} - 1 - \varepsilon \leq \sigma _i  ( \mathbf{\Psi} )   \}$ and $B :=  \{  \sigma _i  ( \mathbf{\Psi} )  \geq \sqrt{{n}/{m}} + 1 + \varepsilon \}$, and (b) is due to Lemma~\ref{sigma of Gauss matrix}. 

\item  {\bf Lower bound of $ \Pr  ( E_2 ) $:} 

From $ \Pr  ( E_1 ) $,  we can also derive a lower bound for $\Pr ( E_2 )$. Specifically,  by the basic probability theory,
\begin{eqnarray} 
 \Pr ( E_2 ) & \hspace{-2mm}  \geq & \hspace{-2mm}  \Pr ( E_2 E_1)  = \Pr (  E_2 | E_1 ) \Pr (E_1)  \nonumber \\
& \hspace{-2mm} \overset{ (a) } {=} & \hspace{-2mm} \Pr (E_1)   \nonumber\\
& \hspace{-2mm} \overset{\eqref{eq:72ygygy}}{\geq} & \hspace{-2mm}  1 - 2 e^{ - { m \varepsilon ^2 }/{2} },  
\label{eq:A73}
\end{eqnarray} 
where (a) is because 
\begin{eqnarray}
\nu_m & \hspace{-2mm} {=} & \hspace{-2mm}  \frac{\sigma_1 ^2  ( \mathbf{\Psi} )   - \sigma_m^2}{\sigma_1^2  ( \mathbf{\Psi} ) + \sigma_m^2} \nonumber \\
 & \hspace{-2mm} \overset{E_1}{ \leq } & \hspace{-2mm}  \frac{  (   \sqrt{{n}/{m}} + 1 +  \varepsilon )^2 -  ( \sqrt{{n}/{m}} - 1 - \varepsilon )^2} { (   \sqrt{{n}/{m}} + 1 +  \varepsilon )^2  + ( \sqrt{{n}/{m}} - 1 - \varepsilon )^2} \nonumber \\
 & \hspace{-2mm}  = & \hspace{-2mm}   \frac{ 2 \sqrt{{n}/{m}} (  1 + \varepsilon )} { {n}/{m} + ( 1 + \varepsilon )^2 }   \label{eq:etam}
\end{eqnarray}
so that $\Pr (  E_2 | E_1 ) = 1$. In other words,  event $E_2$ always happens if $E_1$ holds. 
By letting
\begin{equation}
 \varepsilon = \left ( \frac{ 2 + \sqrt{ 4  -  \eta^ 2}} {\eta } \right ) \sqrt{ \frac{n}{m} }   -1,  \label{eq:iiget}
\end{equation}
we have
\begin{equation}
\frac{ 2 \sqrt{{n}/{m}} (  1 + \varepsilon )} { {n}/{m} + ( 1 + \varepsilon )^2 } = \frac{\eta}{2}. \label{eq:iiiii}
\end{equation}
Then, it follows that
\begin{eqnarray}
 \Pr ( E_2 )  &\hspace{-2mm}\overset{ \eqref{eq:iqivhg}}{ = } & \hspace{-2mm}  \Pr    \bigg \{   \nu_m    \leq    \frac{ 2 \sqrt{{n}/{m}}  (  1 +   \varepsilon )} { {n}/{m} +  ( 1 + \varepsilon )^2 }   \bigg \}   \nonumber \\
 &\hspace{-2mm}  \overset{\eqref{eq:iiiii}}{=}& \hspace{-2mm}  \Pr  \left ( \nu_m  \leq  \frac{\eta}{2}  \right ) 
   \nonumber \\
 &\hspace{-2mm}  \overset{\eqref{eq:iiget}}{\geq}& \hspace{-2mm}  1  -   2 e^{   -  { m } \big(  ( 2 + \sqrt{ 4 -  \eta^ 2} ) \sqrt{ {n}/{m} } /  { \eta }  - 1 \big)^2   /2 }.~~~~~~~
\label{eq:etam2}
\end{eqnarray}

\item  {\bf Lower bound of $ \Pr( \mu \left( \mathbf{\Psi} \right)  \leq \eta - \nu_m)$:}

Having $\Pr (E_2)$ in hand, we are now ready to derive a lower bound for $ \Pr( \mu \left( \mathbf{\Psi} \right)  \leq \eta - \nu_m)$. 
To be specific, 
\begin{eqnarray} 
\lefteqn{ \Pr( \mu \left( \mathbf{\Psi} \right)  \leq \eta - \nu_m) } \nonumber \\  
& \hspace{-2mm} \overset{(a)}{\geq} & \hspace{-2mm} \Pr \left ( \mu \left( \mathbf{\Psi} \right)  \leq \eta - \nu_m \Big | \nu_m \leq \frac{\eta}{2} \right ) \Pr \left ( \nu_m \leq \frac{\eta}{2} \right )  \nonumber \\
&  \hspace{-2mm} \overset{\eqref{eq:etam2}}{\geq} & \hspace{-2mm} \Pr \left ( \mu \left( \mathbf{\Psi} \right)  \leq \frac{\eta}{2}  \Big | \nu_m \leq \frac{\eta}{2}  \right ) \nonumber\\
&  \hspace{-2mm} & \hspace{-2mm}    \times  \Big (  1  -   2 e^{   -  { m } \big( ( 2 + \sqrt{ 4 -  \eta^ 2} ) \sqrt{ {n}/{m} } /  { \eta }  - 1 \big)^2   /2 }  \Big ),~~~~~~
\label{eq:A7}
\end{eqnarray} 
where (a) follows from the fact that $\Pr(A) \geq  \Pr(A B) = \Pr(A|B)  \Pr(B)$  with events $A := \{ \mu \left( \mathbf{\Psi} \right)  \leq \eta - \nu_m \}$ and $B :=  \{  \nu_m \leq \tau \eta \}$. 
Thus, our remaining task is to establish a lower bound of the term $\Pr( \mu \left( \mathbf{\Psi} \right)  \leq ( 1 - \tau ) \eta | \nu_m \leq \frac{\eta}{2} )$ in the right-hand side of~\eqref{eq:A7}. In fact,  observe that  
\begin{eqnarray}
\lefteqn{  \Pr \left ( \mu \left( \mathbf{\Psi} \right)  \leq ( 1 - \tau ) \eta \Big | \nu_m \leq \frac{\eta}{2}  \right ) } \nonumber \\
& \hspace{-2mm}  = & \hspace{-2mm}  
1 -  \Pr \left ( \mu \left( \mathbf{\Psi} \right)  >  \frac{\eta}{2} \Big | \nu_m \leq \frac{\eta}{2}  \right ) \nonumber\\
& \hspace{-2mm} \geq & \hspace{-2mm}  
1 -  \Pr \left ( \mu \left( \mathbf{\Psi} \right)  \geq  \frac{\eta}{2}  \right ) \nonumber\\
& \hspace{-2mm} \overset{ \text{Lemma}~\ref{mutual coherence of Gauss matrix}}{\geq}  & \hspace{-2mm} 1 -  n (n - 1)   \left (e^{ - {m  \eta^2}/{ ( 64 + 8 \eta ) } }  +  e^ { - {m} / {16} } \right)  \nonumber\\
& \hspace{-2mm} \overset{ (a) }{>} &\hspace{-2mm} 1 - 2n(n-1) e^{ - {m  \eta^2}/{ ( 64 + 8 \eta ) } } ,  ~~~~~~
\label{probability: mu of gaussmatrix1} 
\end{eqnarray} 
where (a) is follows from that $\frac{ \eta^2 }{ 64 + 8 \eta} < \frac{1}{16}$ for $\eta \in (0, 1)$. 
Using \eqref{eq:A7} and \eqref{probability: mu of gaussmatrix1}, we have 
\begin{eqnarray} 
\lefteqn{\Pr( \mu \left( \mathbf{\Psi} \right)  \leq \eta - \nu_m) } \nonumber \\
& \geq & \hspace{-2mm} \left (  1 - 2n(n-1) e^{ - {m  \eta^2}/{ ( 64 + 8 \eta ) } }  \right)    \nonumber \\
&  & \hspace{-2mm} \times \left ( 1  -   2 e^{   -  { m } \big(  ( 2 + \sqrt{ 4 -  \eta^ 2} ) \sqrt{ {n}/{m} } /  { \eta }  - 1 \big)^2   /2 }  \right )  \nonumber \\
& > & \hspace{-2mm} 1 - 2n(n-1) e^{ - {m  \eta^2}/{ ( 64 + 8 \eta ) } } \nonumber \\
&  & \hspace{-2mm}  -   2 e^{   -  { m } \big(  ( 2 + \sqrt{ 4 -  \eta^ 2} ) \sqrt{ {n}/{m} } /  { \eta }  - 1 \big)^2   /2 } \nonumber \\
& \overset{ (a) }{ > } & \hspace{-2mm} 1 - 3 n ( n - 1)  e^{ - {m  \eta^2}/{ ( 64 + 8 \eta ) } },  \nonumber\\
& \overset{ (b) }{ > } & \hspace{-2mm} 1 - 3 n^2  e^{ -  {m \eta^2}/{ 72} }, ~~~~~\label{eq:A10}
\end{eqnarray} 
where  (a)  holds true under 
\begin{equation}
 \sqrt{\frac{n}{m}} \geq \left( 1  + \frac{ \eta }{ 2 \sqrt{ 8 + \eta } } \right) \frac{ \eta }{ 2 + \sqrt{ 4 - \eta^2 }} \label{eq:81} 
\end{equation}
and (b) due to $ n ( n - 1) < n^2$ and $ \eta \in ( 0 , 1)$.
\end{enumerate}

Finally, one can verify that the right-hand side of~\eqref{eq:81} increases  monotonically in $\eta \in (0,1)$ and thus is upper bounded by $ 7/(6 ( 2 + \sqrt{3}))  \approx 0.3126$. Therefore, it can be concluded that \eqref{eq:81} is guaranteed whenever $\frac{n}{m} \geq 0.1$. This completes the proof. 
\end{IEEEproof}

\section{Proof of Theorem~\ref{theorem: mols}}\label{app:2}
\subsection{Main Idea}

\begin{IEEEproof}
Our proof   works by mathematical induction, which follows a similar strategy as in~\cite{wang2012Generalized,wang2017MOLS} but with extension to the mutual coherence framework. Assume that the mOLS algorithm  has performed $k$ ($0 \leq k < K$) iterations successfully, that is, it picked up at least one correct index in each of these iterations. Under this assumption, the previously selected index set $\mathcal{S}^k$ contains at least $k$ correct ones ($| \mathcal{S}^k \cap T | \geq k$). We shall show that the algorithm will also perform successfully in the ($k + 1$)-th iteration. 

We first introduce a simple principle to determine if mOLS will perform successfully in the ($k + 1$)-th iteration. Let $U^k = \{ u_1, \cdots, u_s \} \subseteq \Omega \backslash T$ be an index set such that
\begin{equation}
\frac{  | \langle \phi_{u_1} , \mathbf{r}^k \rangle | } { \| \mathcal{P}_{ \mathcal{S}^k }^\bot  \phi_{u_1} \|_2 }  \geq \frac{  | \langle \phi_{u_2} , \mathbf{r}^k \rangle | } { \| \mathcal{P}_{ \mathcal{S}^k }^\bot  \phi_{u_2} \|_2 }  \geq \cdots \geq \frac{  | \langle \phi_{u_s}, \mathbf{r}^k \rangle | } { \| \mathcal{P}_{ \mathcal{S}^k }^\bot  \phi_{u_s} \|_2 }, 
\end{equation}
 where $\phi_i$ is the $i$-th column of $\mathbf{\Phi}$, $\mathcal{P}_{ \mathcal{S}^k }^\bot$ is the orthogonal projection against the subspace spanned by the columns of $\mathbf{\Phi}_{ \mathcal{S}^k }$, and $\mathbf{r}^k$ is the residual vector at the $k$-th iteration of mOLS.  
Then, we define two quantities:
    \begin{equation}
    \alpha_s: = \min_{ i \in U^k} \frac{  | \langle \phi_{i}, \mathbf{r}^k \rangle | } { \| \mathcal{P}_{ \mathcal{S}^k }^\bot  \phi_{i} \|_2 } \label{eq:B.2}
   ~~\text{and}~~
    \beta _1 := \max_{ i \in T \backslash \mathcal{S}^k} \frac{ |  \langle \phi_i,  \mathbf{r}^k \rangle | } {~\| \mathcal{P}_{\mathcal{S}^k} ^\bot \mathbf{\phi} _i \|_2 } 
    \end{equation}

By the selection rule of the mOLS algorithm,  it will perform successfully in the $(k+1)$-th iteration whenever
\begin{equation}
\beta _1 > \alpha _s.   \label{sufficient condition}
\end{equation} 
In the following, we shall build an upper bound for $\alpha_s$ and a lower bound for $\beta_1$, respectively. Combining these two bounds leads to a sufficient condition for the algorithm to be successful in the $(k+1)$-th iteration.

\subsection{Lower Bound of $\beta_1$}
We take some observations on the residual $\mathbf{r}^k$. From~\eqref{eq:rrrrr}, we can formulate the residual $\mathbf{r}^k $ as
%
    \begin{eqnarray}
      \mathbf{r}^k &\hspace{-2mm}=&\hspace{-2mm}  \mathcal{P}_{ \mathcal{S}^k }^\bot \mathbf{y}. 
      =  \mathcal{P}_{\mathcal{S}^k}^ \bot  \mathbf{\Phi }_T \mathbf{x}_T \nonumber \\
      &\hspace{-2mm}=&\hspace{-2mm}  \mathcal{P}_{\mathcal{S}^k}^\bot  ( \mathbf{\Phi }_{\mathcal{S}^k} \mathbf{x}_{\mathcal{S}^k} + \mathbf{\Phi }_{T \backslash \mathcal{S}^k} \mathbf{x}_{T \backslash \mathcal{S}^k} ) = \mathcal{P}_{\mathcal{S}^k}^\bot \mathbf{\Phi }_{T \backslash \mathcal{S}^k} \mathbf{x}_{T \backslash \mathcal{S}^k},~~~~~~  \label{eq:resi_k}
    \end{eqnarray}
    where the last equality is because the orthogonal projector $\mathcal{P}_{\mathcal{S}^k}^\bot$ eliminates all the components lying in the subspace spanned by the columns of $\mathbf{\Phi }_{\mathcal{S}^k}$. 

Using \eqref{eq:resi_k}, $\beta_1$ can be lower bounded as follows, 
\begin{eqnarray}
  \beta _1 & \hspace{-4mm} =&\hspace{-3mm} \mathop {\max } \limits_{i \in T \backslash \mathcal{S}^k} \frac{ | \langle \phi_i, \mathbf{r}^k  \rangle | } {~ \| \mathcal{P}_{\mathcal{S}^k}^ \bot \phi _i \|_2} \nonumber \\
 &\hspace{-4mm}\stackrel{(a)} \geq &\hspace{-3mm} 
 \| \mathbf{\Phi} _{T \backslash \mathcal{S}^k}^T \mathbf{r}^k \|_\infty \nonumber \\
&\hspace{-4mm} \stackrel{(b)} \geq &\hspace{-3mm} \frac{ \| \mathbf{\Phi} _{T \backslash \mathcal{S}^k}^T \mathbf{r}^k  \|_2} {\sqrt{\left| T\backslash \mathcal{S}^k \right|}} \nonumber \\
 &\hspace{-4mm} \overset{\eqref{eq:resi_k}} {=} &\hspace{-3mm} \frac{ \| \mathbf{\Phi} _{T \backslash \mathcal{S}^k}^T  \mathcal{P}_{\mathcal{S}^k}^\bot \mathbf{\Phi }_{T \backslash \mathcal{S}^k} \mathbf{x}_{T \backslash \mathcal{S}^k} \|_2 }{\sqrt{\left| T\backslash \mathcal{S}^k \right|}}  \nonumber\\
 &\hspace{-4mm}\stackrel{(c)}{\geq}& \hspace{-3mm} \frac{\| \mathbf{\Phi}_{T \backslash \mathcal{S}^k}^T   \hspace{-.25mm}  \mathbf{\Phi} _{T \backslash \mathcal{S}^k} \mathbf{x}_{T\backslash \mathcal{S}^k}   \hspace{-.25mm}  \|_2  \hspace{-.5mm}  -  \hspace{-.5mm}  \| \mathbf{\Phi} _{T\backslash \mathcal{S}^k}^T   \hspace{-.25mm}  \mathcal{P}_{\mathcal{S}^k}   \hspace{-.25mm}  \mathbf{\Phi} _{T \backslash \mathcal{S}^k} \mathbf{x}_{T \backslash \mathcal{S}^k}  \hspace{-.25mm}   \|_2}{\sqrt{ | T\backslash \mathcal{S}^k |}}, \nonumber \\ \label{eq:b1}
\end{eqnarray}
where (a) is because $\| \phi_i \|_2 = 1$ and thus ${{\left\| {\mathcal{P}_{{\mathcal{S}^k}}^ \bot {\mathbf{\phi} _i}} \right\|}_2}\le 1$, (b) is due to Lemma \ref{norminquality} and (c) is from $ \mathcal{P}_{\mathcal{S}^k}^\bot = \mathbf{I} -   \mathcal{P}_{\mathcal{S}^k}$.
In fact, the two terms in the numerator of the right-hand side of~\eqref{eq:b1} can respectively be bounded as 
\begin{eqnarray}\label{eq:b11}
\lefteqn{  \| \mathbf{\Phi} _{T \backslash \mathcal{S}^k}^T \mathbf{\Phi} _{T \backslash \mathcal{S}^k} \mathbf{x}_{T \backslash \mathcal{S}^k} \|_2  } \nonumber \\
 &  \stackrel{\eqref{rip:consequence:1}} {\geq} & \hspace{-2mm} ( 1 - \delta_{|T \backslash \mathcal{S}^k|} )   \| \mathbf{x}_{T \backslash \mathcal{S}^k}  \|_2 \nonumber \\
 &  \stackrel{{\eqref{rip:consequence:3}}} {\geq} & \hspace{-2mm} ( 1  -  ( | T \backslash \mathcal{S}^k |  - 1 ) \mu )  \| \mathbf{x}_{T \backslash \mathcal{S}^k} \|_2 
\end{eqnarray} 
and
\begin{eqnarray}
\lefteqn {  \| \mathbf{\Phi} _{T \backslash \mathcal{S}^k}^{T} \mathcal{P}_{\mathcal{S}^k} \mathbf{\Phi} _{T\backslash \mathcal{S}^k} \mathbf{x}_{T \backslash \mathcal{S}^k} \|_2 } \nonumber \\
& = & \hspace{-2mm} \| \mathbf{\Phi} _{T \backslash \mathcal{S}^k}^{T} \mathbf{\Phi} _{ \mathcal{S}^k} ( \mathbf{\Phi} _{ \mathcal{S}^k}^T \mathbf{\Phi} _{\mathcal{S}^k} )^{- 1}  \mathbf{\Phi} _{\mathcal{S}^k}^T \mathbf{\Phi} _{T \backslash \mathcal{S}^k} \mathbf{x}_{T \backslash \mathcal{S}^k} \|_2  \nonumber \\
 & \stackrel{(a)} {\leq} & \hspace{-2mm} \mu \sqrt { | T \backslash \mathcal{S}^k | \cdot | \mathcal{S}^k | } \|  ( \mathbf{\Phi} _{\mathcal{S}^k}^{T} \mathbf{\Phi} _{ \mathcal{S}^k} )^{ - 1 } \mathbf{\Phi} _{\mathcal{S}^k}^T \mathbf{\Phi} _{T \backslash \mathcal{S}^k} \mathbf{x}_{T \backslash \mathcal{S}^k} \|_2 \nonumber \\
 &  \stackrel{(b)}{\leq} & \hspace{-2mm} \frac{\mu \sqrt { | T \backslash \mathcal{S}^k | \cdot | \mathcal{S}^k |}} { 1 -  (  | \mathcal{S}^k  | - 1  )  \mu } \| \mathbf{\Phi} _{\mathcal{S}^k}^{T} \mathbf{\Phi} _{T \backslash \mathcal{S}^k} \mathbf{x}_{T\backslash \mathcal{S}^k} \|_2 \nonumber \\
 & \stackrel{(c)}{\leq} & \hspace{-2mm}  \frac{\mu    | T \backslash \mathcal{S}^k  | \cdot | \mathcal{S}^k |   } {1 - ( | \mathcal{S}^k  | - 1 ) \mu } \| \mathbf{x}_{T \backslash \mathcal{S}^k} \|_2,
\label{eq:b12}
\end{eqnarray}
where (a) and (c) follow from Lemma \ref{lem:mip_1}, and (b) is due to~\eqref{rip:consequence:2} and \eqref{rip:consequence:3}.

Therefore, by  combining \eqref{eq:b11} and \eqref{eq:b12}, we have
\begin{eqnarray}
\hspace{-.125mm}  \beta_1 \hspace{-.25mm} \geq \hspace{-.25mm} \left ( \hspace{-.25mm} 1 \hspace{-.25mm} -  \hspace{-.25mm} (  | T \backslash \mathcal{S}^k |  \hspace{-.25mm} - \hspace{-.25mm} 1 ) \mu  \hspace{-.25mm} - \hspace{-.25mm} \frac{ | T \backslash \mathcal{S}^k | \cdot | \mathcal{S}^k | \mu } {1 \hspace{-.25mm} -  \hspace{-.25mm}  ( | \mathcal{S}^k |  \hspace{-.25mm} -  \hspace{-.25mm} 1 ) \mu } \hspace{-.25mm} \right )   \hspace{-.25mm}
    \frac{\| \mathbf{x}_{T \backslash \mathcal{S}^k} \|_2}{ \sqrt { | T \backslash \mathcal{S}^k  | } }. \label{eq:up} \hspace{-.25mm}
\end{eqnarray}

\subsection{Upper Bound of $\alpha_s$}
From~\eqref{eq:B.2}, 
%
\begin{eqnarray}
\alpha_s  
 &\hspace{-2mm} \overset{ (a) } {\leq}  & \hspace{-2mm} \frac{1}{s} \sum \limits_{ i \in U^k } \frac{ | \langle \phi _i, \mathbf{r}^k \rangle |} { \| \mathcal{P}_{ \mathcal{S}^k }^ \bot \phi_i  \|_2} \nonumber \\
 &\hspace{-2mm} \overset{ (b) } {\leq}  &  \hspace{-2mm} \frac{\| \mathbf{ \Phi } _ { U^k } ^ {T} \mathbf{r}^k \|_1} {s  \mathop {\min } \limits_{ i \in U^k} \| \mathcal{P}_{ \mathcal{S}^k} ^ \bot \phi_i \|_2 }  \nonumber\\
 & \hspace{-2mm} \overset{(c)} { \leq } & \hspace{-2mm}  \frac{ \sqrt{s} \| \mathbf{\Phi}_{U^k}^ {T} \mathbf{r}^k \|_2} {s \mathop {\min } \limits_{ i \in U^k } \| \mathcal{P}_{ \mathcal{S}^k} ^ \bot \phi _i \|_2 }  = \frac{ \| \mathbf{\Phi}_{U^k}^ {T} \mathbf{r}^k \|_2} {\sqrt{s}  \mathop {\min } \limits_{ i \in U^k } \| \mathcal{P}_{ \mathcal{S}^k} ^ \bot \phi _i \|_2 } ,  \label{eq:3}
\end{eqnarray}
where (a) is because the smallest element of a set should be no larger than its average, (b) is due to $\mathop {\min } \limits_{ i \in U^k} \| \mathcal{P}_{ \mathcal{S}^k} ^ \bot \phi_i \|_2 \leq \| \mathcal{P}_{ \mathcal{S}^k} ^ \bot \phi_j \|_2,~\forall{j \in U^k }, $ and (c) follows from Lemma~\ref{norminquality}.

Next, we take some observations on the right-hand side of~\eqref{eq:3}. 
Since matrix $\mathbf{\Phi}$ has unit $\ell_2$-norm columns,  the term $\mathop {\min } \limits_{ i \in U^k } \| \mathcal{P}_{ \mathcal{S}^k} ^ \bot \phi _i \|_2$ in the denominator satisfies 
\begin{eqnarray}
\mathop {\min } \limits_{ i \in U^k } \| \mathcal{P}_{ \mathcal{S}^k} ^ \bot \phi _i \|_2 &\hspace{-2mm}    \overset{ \rm{Lemma}~\ref{lemma:OlS} } {\geq} &\hspace{-2mm}   \sqrt{ 1 - \delta_{| \mathcal{S}^k | + 1 }} \nonumber \\
 &\hspace{-2mm}  \overset{(a)} {\geq} &\hspace{-2mm}   1-\delta_{| \mathcal{S}^k | + 1} \nonumber\\
 &\hspace{-2mm}  \overset{\eqref{rip:consequence:3}} {\geq} &\hspace{-2mm}  1 - | \mathcal{S}^k | \mu, \label{eq:4}
\end{eqnarray}
where (a) holds because of $1-\delta_{ | \mathcal{S}^k | + 1} < 1$.

Building a lower bound for the numerator of the right-hand side of \eqref{eq:3} requires a little more effort.  In fact, 
\begin{eqnarray}
  \lefteqn{ \| \mathbf{\Phi}_{U^k}^{T} \mathbf{r}_k \|_2 } \nonumber \\ 
  & \hspace{-2mm} = & \hspace{-2mm} \| \mathbf{\Phi}_{U^k}^{T} ( \mathbf{I} - \mathcal{P}_{\mathcal{S}^k} ) \mathbf{\Phi} _{T \backslash \mathcal{S}^k} \mathbf{x}_{T \backslash \mathcal{S}^k} \|_2 \nonumber\\
 & \hspace{-2mm} \stackrel{(a)} {\leq} & \hspace{-2mm} \| \mathbf{\Phi}_{U^k}^{T} \mathbf{\Phi}_{ T \backslash \mathcal{S}^k } \mathbf{x}_{ T \backslash \mathcal{S}^k} \|_2   \hspace{-.25mm} +   \hspace{-.25mm} \| \mathbf{\Phi}_{U^k}^{T} \mathcal{P}_{ \mathcal{S}^k} \mathbf{\Phi}_{ T \backslash \mathcal{S}^k } \mathbf{x}_{ T \backslash \mathcal{S}^k } \|_2~~~~~~~~~
 \label{eq:a1}
\end{eqnarray}
where (a) is due to \eqref{norminquality:1}.  The two terms in the right-hand side of \eqref{eq:a1} can, respectively,  be bounded as 
\begin{eqnarray}
 \| \mathbf{\Phi}_{U^k}^{T} \mathbf{\Phi} _{ T \backslash \mathcal{S}^kn} \mathbf{x}_{ T \backslash \mathcal{S}^k } \|_2  & \hspace{-3mm} \overset {\rm{Lemma} ~\ref{lem:mip_1}} {\leq} & \hspace{-3mm} \mu \sqrt{ | U^k |   \hspace{-.25mm}   \cdot  \hspace{-.25mm} | T \backslash \mathcal{S}^k | } \| \mathbf{x}_{ T \backslash \mathcal{S}^k } \|_2 \nonumber\\
&\hspace{-3mm} = & \hspace{-3mm} \mu \sqrt{ s  | T \backslash \mathcal{S}^k | } \| \mathbf{x}_{ T \backslash \mathcal{S}^k} \|_2,   ~~~~~~~~~ \label{eq:a11}
\end{eqnarray}
and
\begin{eqnarray}
\lefteqn{ \| \mathbf{\Phi}_{U^k}^{T} \mathcal{P}_{ \mathcal{S}^k } \mathbf{\Phi} _{ T \backslash \mathcal{S}^k} \mathbf{x}_{ T \backslash \mathcal{S}^k} \|_2 } \nonumber \\
&  = & \hspace{-2mm} \| \mathbf{\Phi}_{U^k}^ {T} \mathbf{\Phi} _{\mathcal{S}^k} ( \mathbf{\Phi}_{ \mathcal{S}^k}^{T} \mathbf{\Phi} _ {\mathcal{S}^k} ) ^{ - 1 } \mathbf{\Phi} _{ \mathcal{S}^k}^{T} \mathbf{\Phi} _{ T \backslash \mathcal{S}^k} \mathbf{x}_{ T \backslash \mathcal{S}^k} \|_2 \nonumber\\
 &\overset{\rm{Lemma}~\ref{lem:mip_1}} {\leq} & \hspace{-2mm} \mu \sqrt{ | U^k | \cdot | \mathcal{S}^k | } \| ( \mathbf{\Phi} _{\mathcal{S}^k}^{T} \mathbf{\Phi}_{ \mathcal{S}^k} )^{ - 1 } \mathbf{\Phi} _{\mathcal{S}^k}^{T} \mathbf{\Phi}_{ T \backslash \mathcal{S}^k } \mathbf{x}_{ T \backslash \mathcal{S}^k } \|_2 \nonumber\\
 & \overset{(a)} {\leq} & \hspace{-2mm} \frac{\mu \sqrt { s  | \mathcal{S}^k |}  \| \mathbf{\Phi} _{ \mathcal{S}^k}^{T} \mathbf{\Phi} _{ T \backslash \mathcal{S}^k} \mathbf{x}_{ T \backslash \mathcal{S}^k} \|_2} { 1 - ( | \mathcal{S}^k | - 1 ) \mu }  \nonumber\\
 & \overset{\rm{Lemma}~\ref{lem:mip_1}} {\leq} & \hspace{-2mm} \frac{ \mu \sqrt { s | \mathcal{S}^k |}} { 1 - ( | \mathcal{S}^k | - 1 ) \mu } \left( \mu \sqrt{ | T \backslash \mathcal{S}^k | \cdot | \mathcal{S}^k |} \| \mathbf{x}_{ T \backslash \mathcal{S}^k}\|_2   \right) \nonumber\\
 & = & \hspace{-2mm} \frac{ | \mathcal{S}^k | \sqrt{ s | T \backslash \mathcal{S}^k |} \mu ^2} { 1 - ( | \mathcal{S}^k | - 1 ) \mu } \| \mathbf{x}_{ T \backslash \mathcal{S}^k } \|_2, \label{eq:a12}
\end{eqnarray}
where (a) follows from \eqref{rip:consequence:3} and \eqref{rip:consequence:2}.

Using \eqref{eq:a1},  \eqref{eq:a11}) and \eqref{eq:a12}, we get
\begin{eqnarray}
\lefteqn{ \| \mathbf{\Phi}_{U^k}^{T} \mathbf{r}^k \|_2 } \nonumber \\
&  \leq &  \hspace{-3mm}  \left(  \hspace{-.25mm} \mu \sqrt{ s | T \backslash \mathcal{S}^k | } + \frac{ | \mathcal{S}^k | \sqrt{ s  | T \backslash \mathcal{S}^k | } \mu ^2 } { 1 -  ( | \mathcal{S}^k |  -  1 ) \mu }  \right)  \|\mathbf{x}_{ T \backslash \mathcal{S}^k } \|_2. ~~~~~~~~~\label{eq:s0}
\end{eqnarray}

Finally, plugging \eqref{eq:4} and \eqref{eq:s0} into \eqref{eq:3}, we have
\begin{equation}
\hspace{-0.25mm}  \alpha_s  \leq   \frac{ \sqrt{ | T \backslash \mathcal{S}^k |}} { 1  -  | \mathcal{S}^k | \mu } \left( \mu   +  \frac{ | \mathcal{S}^k | \mu ^2} { 1 - ( | \mathcal{S}^k |   -  1 ) \mu } \right ) \| \mathbf{x}_{ T \backslash \mathcal{S}^k } \|_2.  \hspace{-0.25mm}  \label{eq:down} 
\end{equation}

\subsection{Overall Condition}
Thus far, we have established in~\eqref{eq:down} an upper bound for $\alpha_s$  and  in \eqref{eq:up} a lower bound for $\beta_1$. By relating these two bounds, we have that $\beta _1 > \alpha _s$ can be satisfied when
\begin{eqnarray}
 \lefteqn{ \frac{1} { \sqrt { | T \backslash \mathcal{S}^k |}  } \left( 1 - ( | T \backslash \mathcal{S}^k | - 1 ) \mu  - \frac{ | T \backslash \mathcal{S}^k | \cdot | \mathcal{S}^k | \mu ^2} { 1 - ( | \mathcal{S}^k | - 1 ) \mu } \right ) }\nonumber\\
 &~~~~~~~~~~ > & \hspace{-2mm} \frac{ \sqrt { | T \backslash \mathcal{S}^k |}} { 1 - |\mathcal{S}^k| \mu } \left( \mu + \frac{ | \mathcal{S}^k | \mu ^2 } { 1 - ( | \mathcal{S}^k | - 1 ) \mu } \right ), ~~~~~~~\label{sufficient_condition:1}
\end{eqnarray}
which is equivalent to
\begin{equation}
\frac{1}  { | T \backslash \mathcal{S}^k | } > { \frac{\mu } { 1 - ( | \mathcal{S}^k | - 1 ) \mu}  + \frac{ \mu } { ( 1 - ( | \mathcal{S}^k | - 1) \mu ) ( 1- | \mathcal{S}^k |\mu)}}. \label{eq:c19}
\end{equation}

Since $$ \frac{\mu } { 1 - ( | \mathcal{S}^k | - 1 ) \mu} < \frac{ \mu } { ( 1 - ( | \mathcal{S}^k | - 1) \mu ) ( 1- | \mathcal{S}^k |\mu)},$$
it is clear that \eqref{eq:c19} holds under
\begin{equation}
  \frac{1}{| T \backslash \mathcal{S}^k |}   >  {\frac{2\mu }{ ( 1 - ( | \mathcal{S}^k | - 1) \mu ) ( 1- | \mathcal{S}^k |\mu)}}, 
\end{equation}
or equivalently, 
\begin{eqnarray}
 {| T \backslash \mathcal{S}^k |}  & \hspace{-2mm} < & \hspace{-2mm}  {\frac{ ( 1 - ( | \mathcal{S}^k | - 1) \mu ) ( 1- | \mathcal{S}^k |\mu)}{2\mu }} \nonumber \\
 & \hspace{-2mm} = & \hspace{-2mm} \frac{1}{2}\left(  \frac{1} {\mu } + 1 \right) - | \mathcal{S}^k | + \frac{| \mathcal{S}^k  | (| \mathcal{S}^k | -1 ) \mu} {2}. ~~~~~ \label{eq:C21}
\end{eqnarray}

Furthermore, since the third term $\frac{| \mathcal{S}^k  | (| \mathcal{S}^k | -1 ) \mu} {2}$ in the right-hand side of~\eqref{eq:C21} is always non-negative for $| \mathcal{S}^k | = s k $ and $k \in \{ 0, 1, \cdots, K - 1 \}$, \eqref{eq:C21} is ensured whenever
\begin{equation}
 {| T \backslash \mathcal{S}^k |} < \frac{1}{2}\left(  \frac{1} {\mu } + 1 \right) - | \mathcal{S}^k |.  \label{eq:22}
\end{equation}
By induction hypothesis the previously selected index set $\mathcal{S}^k$ contains at least $k$ correct ones ($| \mathcal{S}^k \cap T | \geq k$), which implies that $$\left| T \backslash \mathcal{S}^k \right| \leq K - k,$$ and hence \eqref{eq:22} is guaranteed under  
\begin{equation}
K - k < \frac{1}{2} \left( 1 + \frac{1}{\mu } \right) - s k.  \label{eq:sufficient condition1}
\end{equation}

Noting that $ 0\leq k \leq K-1$, \eqref{eq:sufficient condition1} holds if
\begin{equation}
    s K - s +1 < \frac{1} {2} \left( 1 + \frac{1}{\mu } \right),
\end{equation}
that is
\begin{equation} \label{sufficient:s=1}
    \mu < \frac{1} {2sK - 2s + 1}. 
\end{equation}

In summary,  under \eqref{sufficient:s=1} the mOLS will select all support indices within $K$ iterations. Let $k^*$ denote the number of iterations when mOLS stops. Then, the estimated support set $T^{k^*}$ may contain some indices that do not belong to $T$. Even in this situation, the final recovery result is
unaffected  (i.e., $\hat{\mathbf{x}} = {\mathbf{x}}^{k^*}
= \mathbf{x}$) because 
\begin{eqnarray} \label{eq:LS}
 {\mathbf{x}}^{k^*} = \underset{\mathbf{u}: {supp}(\mathbf{u}) = T^{k^*}}{ \arg \min} {\|\mathbf{y}-\mathbf{\Phi}\mathbf{u}\|}_{2} \nonumber
\end{eqnarray}
and
\begin{eqnarray}
({\mathbf{x}}^{k^*})_{T^{k^*}} &=& \label{eq:bff1} \mathbf{\Phi}^\dag_{T^{k^*}}  \mathbf{y} ~=~ \label{eq:bff1} \mathbf{\Phi}^\dag_{T^{k^*}}  \mathbf{\Phi}_{T} \mathbf{x}_{T} \nonumber
 \\
 &\overset{(a)}{=}& \mathbf{\Phi}^\dag_{T^{k^*}} \mathbf{\Phi}_{T^{k^*}} \mathbf{x}_{T^{k^*}} = \label{eq:bff3} \mathbf{x}_{T^{k^*}},
\end{eqnarray}
where (a) is due to the fact that  $\text{supp}(\mathbf{x}) = T$ and thus
$\mathbf{x}_{T^{k^*} \backslash T} = \mathbf{0}$,
which completes the proof. 
\end{IEEEproof}

\section{Proof of theorem \ref{theorem:pmols}}\label{app:3}
\begin{IEEEproof}
We cannot directly combine Theorem~\ref{theorem: mu of preconditioned matrix} and Theorem~\ref{theorem: mols} because the latter requires the sampling matrix $\mathbf{\Phi}$ to have unit $\ell_2$-norm columns. However, if $\mathbf{\Psi}\in \mathbb{R}^{m\times n}$ has entries ${\varphi _{ij}}\mathop  \sim \limits^{i.i.d} \mathcal{N}  (0,{1}/{m} )$, then $\mathbf{\Phi} =  \mathbf{P}\mathbf{\Psi}$ cannot always satisfy the unit $\ell_2$-norm constraint. Nevertheless, we can address this issue by reformulating the sampling system. To be specific, 
\begin{eqnarray}
\mathbf{y} & \hspace{-2mm} = & \hspace{-2mm} \mathbf{\Phi} \mathbf{x} \nonumber \\
& \hspace{-2mm} = & \hspace{-2mm} 
\underbrace{ \left [ \frac{\mathbf{\Phi}_1}{\|\mathbf{\Phi}_1\|_2}, \cdots, \frac{\mathbf{\Phi}_n}{\|\mathbf{\Phi}_n\|_2} \right ]}_{\bar{\mathbf{\Phi}}}
\hspace{-.5mm}
\underbrace{ \left[ \begin{array}{*{20}{c}}
{\|\mathbf{\Phi}_1\|_2}&  & \mathbf{0}  \\
 & \hspace{-1mm} \ddots   \hspace{-1mm} &  \\
\mathbf{0}  &  &  {\|\mathbf{\Phi}_n\|_2}
\end{array} \right]  \hspace{-1mm}
  \left[ \begin{array}{c}
      \\
    \hspace{-1mm}  \mathbf{x}  \hspace{-1mm}  \\
    \\
    \end{array} \right]  }_{\bar{\mathbf{x}}}, \nonumber \\
\end{eqnarray}
where matrix $\bar{\mathbf{\Phi}}$ has unit $\ell_2$-norm columns and $\bar{\mathbf{x}}$ is $K$-sparse. 

According to Theorem~\ref{theorem: mols}, the mOLS algorithm exactly recovers any $K$-sparse signal $\bar{\mathbf{x}}$ from it samples $\mathbf{y} = \bar{\mathbf{\Phi}} \bar{\mathbf{x}}$
if 
\begin{equation}
\mu(\bar{\mathbf{\Phi}})  < \frac{1} {2sK - 2s + 1}
\end{equation} 
Since $\mu \left(  \bar{\mathbf{\Phi} } \right) = \mu (  \mathbf{\Phi} ) = \mu( \mathbf{P} \mathbf{\Psi})$ by definition of mutual coherence in \eqref{mutual coherence}, and also noting that $\mathbf{x}$ immediately follows from $\bar{\mathbf{x}}$, i.e., 
 \begin{equation}
\mathbf{x} = \left[ \begin{array}{*{20}{c}}
\frac{1}{\|\mathbf{\Phi}_1\|_2}&  & \mathbf{0}  \\
 & \ddots&  \\
\mathbf{0}  &  &  \frac{1}{\|\mathbf{\Phi}_n\|_2}
\end{array} \right]  \hspace{-1mm}
  \left[ \begin{array}{c}
      \\
    \hspace{-1mm}  \bar{\mathbf{x}}  \hspace{-1mm}  \\
    \\
    \end{array} \right], 
 \end{equation}
 we can conclude that the mOLS algorithm exactly recovers any $K$-sparse signal $\mathbf{x}$ under 
 \begin{equation}
     \mu ( \mathbf{P} \mathbf{\Psi} ) < \frac{1} {2sK - 2s + 1}. 
 \end{equation}
 This, together with Theorem~\ref{theorem: mu of preconditioned matrix}, implies Theorem~\ref{theorem:pmols}. 
\end{IEEEproof}

\end{document}